\RequirePackage{ifpdf}
\documentclass[11pt,a4paper]{article}
\usepackage{epsfig,jheppubme}

\usepackage{multirow,longtable,enumerate}

\usepackage{physics}
\usepackage{tensor}
\usepackage{amsmath}
\usepackage{amssymb}
\usepackage{appendix}
\usepackage{mathtools}
\usepackage{mathrsfs}
\usepackage[normalem]{ulem}
\usepackage{tikz}
\usepackage{booktabs}

\font\cmss=cmss12

\newcommand\half{\frac12}

\newcommand\bi{\begin{itemize}}
\newcommand\ei{\end{itemize}}

\newcommand\mS{\mathcal{S}}
\newcommand\hmS{{\hat\mS}}

\newcommand\hJ{{\hat J}}
\newcommand\hS{{\hat S}}

\newcommand\bea{\begin{eqnarray}}
\newcommand\eea{\end{eqnarray}}
\newcommand\be{\begin{equation}}
\newcommand\ee{\end{equation}}
\newcommand\nn{\nonumber}

\newcommand\cO{{\cal O}}

\newcommand\cN{{\cal N}}

\newcommand\sfrac[2]{{\textstyle\frac{#1}{#2}}}

\newcommand\shalf{{\textstyle\frac12}}

\newcommand\ZZ{\hbox{Z\kern-.4emZ}}
\newcommand\sZZ{\hbox{\sevenfont Z\kern-.4emZ}}

\newcommand{\eref}[1]{Eq.\,(\ref{#1})}
\newcommand{\Comment}[1]{{}}

\newcommand{\mc}{\mathcal}
\newcommand{\msc}{\mathscr}
\newcommand{\mf}{\mathfrak}

\newcommand{\mfs}{\mathfrak s}
\newcommand{\mfc}{\mathfrak c}
\allowdisplaybreaks

\def\IB{\relax{\rm I\kern-.18em B}}
\def\IC{{\relax\hbox{\kern.3em{\cmss I}$\kern-.4em{\rm C}$}}}
\def\ID{\relax{\rm I\kern-.18em D}}
\def\IE{\relax{\rm I\kern-.18em E}}
\def\IF{\relax{\rm I\kern-.18em F}}
\def\II{\relax{\rm I\kern-.18em I}}

\def\Id{\relax{1\kern-.32em 1}}
\def\IG{\relax\hbox{$\inbar\kern-.3em{\rm G}$}}
\def\IR{\relax{\rm I\kern-.18em R}}

\renewcommand{\Im}{{\rm Im\,}}

\title{Contour Integrals and the Modular $\mS$-Matrix} \author{Sunil Mukhi \footnote{Email: sunil.mukhi@gmail.com}, Rahul Poddar \footnote{Email: rahul.poddar.305@gmail.com} and Palash Singh \footnote{Email:
palash13singh@gmail.com}\\ \it Indian Institute of Science Education and Research,\\ \it Homi Bhabha Rd, Pashan, Pune 411 008, India
} 

\abstract{We investigate a conjecture to describe the characters of large families of RCFT's in terms of contour integrals of Feigin-Fuchs type. We provide a simple algorithm to 
determine the modular $\mS$-matrix for arbitrary numbers of characters as a sum over paths. Thereafter we focus on the case of 2, 3 and 4 characters, where agreement between the critical exponents of the integrals and the characters implies that the conjecture is true. In these cases, we compute the modular $S$-matrix explicitly, verify that it agrees with expectations for known theories, and use it to compute degeneracies and multiplicities of primaries. We also compute $\mS$ in an 8-character example to provide additional evidence for the original conjecture. On the way we note that the Verlinde formula provides interesting constraints on the critical exponents of RCFT in this context.}

\preprint{}


\arxivnumber{}

\begin{document}

\maketitle

\section{Introduction}

\label{intro}

A procedure for the classification of admissible characters for rational conformal field theories in 2 dimensions was proposed in \cite{Mathur:1988na}. This starts by writing the most general Modular Linear Differential Equation (MLDE) for a given class of theories, labelled by the number $n$ of characters and the Wronskian index $\ell$. This MLDE depends on a finite number of parameters. The $n$ independent solutions automatically transform as vector-valued modular functions, but in general they do not have integral coefficients in their expansion in powers of the parameter $q=e^{2\pi i\tau}$. Thus they cannot be interpreted as counting the degeneracy of states in a physical system. One then varies the parameters in the MLDE until these coefficients become non-negative integers. At this point one has admissible characters and can try to identify the RCFT that they potentially describe. A recent status report on this programme can be found in \cite{Mukhi:2019xjy}.

One deficiency of the MLDE approach is that,  because we solve the equation as a $q$-series, it is difficult or impossible to actually compute the modular transformation matrix $\mS_{ij}$ on the characters. Computations of this matrix rely on being able to express the characters in terms of suitable special functions. This was done long ago for $c<1$ minimal models and WZW models, for which the $\theta$-function representation provides the desired answer via Poisson resummation \cite{Gepner:1986wi, Gepner:1986ip, Cappelli:1986hf}. It was also done in \cite{Naculich:1988xv, Mathur:1988gt, Chandra:2018pjq} for two-character theories by transforming the MLDE into the hypergeometric equation and using the well-known monodromy transformations of hypergeometric functions. Finally, in \cite{Mathur:1988gt} this was achieved for three-character theories with vanishing Wronskian index $\ell$. Here the key observation was that the corresponding MLDE is solved by a set of three contour integrals of Feigin-Fuchs type, having the same integrand but different integration contours. This representation allows us to explicitly compute the modular transformations by rearranging contours and changing variables. In \cite{Mathur:1988gt} this played an important role in the classification programme for the three-character, $\ell=0$ case (for more recent work on this question see \cite{Franc:2019} as well as comments in \cite{Mukhi:2019xjy}).

In \cite{Mukhi:1989qk}, building on the three-character result of \cite{Mathur:1988gt}, it was conjectured that  a similar contour-integral representation could describe the characters of large classes of RCFT with  {\em arbitrary} numbers of characters, in terms of contour integrals of Feigin-Fuchs type. These integrals had previously been used in \cite{Dotsenko:1984ad, Dotsenko:1984nm} to describe correlation functions in a class of RCFT. The  proposal of \cite{Mukhi:1989qk} adapted the same functions, after specialising the parameters and changing some of the prefactors, to describe characters rather than correlators. These contour integral representations depend on very few parameters, which determine the critical exponents of the putative RCFT. As noted in \cite{Mukhi:1989qk}, there is no {\em a priori} reason why the resulting formulae should ``fit'' the exponents of known theories for general $n$. To start with, one finds this is possible only if the Wronskian index $\ell$ vanishes. So at best, the contour integral representation can only describe $\ell=0$ RCFT. Within this class,  \cite{Mukhi:1989qk} found two very surprising results: (i) the critical exponents of many known theories with $\ell=0$, including all $c<1$ minimal models and infinite families of WZW models, can be reproduced using the contour integral representation, (ii) there is  evidence that in all these cases the contour integral representation correctly reproduces the corresponding characters. 

In the present work we further develop the proposal of \cite{Mathur:1988gt,Mukhi:1989qk}. We re-examine the issue of how to compute the modular $\mS$-matrices from contour integrals and find an elegant and simple prescription to do so. 
On the way, we develop formulae to explicitly evaluate the normalisation of the contour integrals, essential to their identification as characters. This is done using formulae for the Selberg integral. Using these results we then compute degeneracies of primary fields in many examples where they were not otherwise known. We also provide additional evidence for the conjecture that contour integrals describe the characters of certain families of models, and  explore the idea that integrality of fusion rules might provide an approach  to the classification of RCFT's via the Verlinde formula\cite{Verlinde:1988sn}.

The outline of this paper is as follows. We start by reviewing the proposal of \cite{Mathur:1988gt,Mukhi:1989qk} and list known theories whose characters are described by contour integrals. Thereafter we present our general method to evaluate the modular $\mS$-matrix. This consists of two parts: (i) a ``sum over paths'' algorithm to determine a related matrix called $\hmS$, (ii) conversion of $\hmS$ to $\mS$ after calculating the normalisations of the integrals. We verify our method by constructing the normalisations and $\mS$-matrices for 2-, 3- and 4-character cases. Using these results we compute degeneracies and multiplicities of primaries in many examples for the first time. Finally we compute the modular $\mS$-matrix for an 8-character theory and show that it gives the expected answer. We conclude by listing some interesting future directions. In three Appendices we illustrate the ``unfolding'' procedure that is used to compute the $\hmS$-matrix, we exhibit different forms of the relevant contour integrals and their mutual relationships, and we list some degeneracies computed by our method.

\section{Contour-integral representation of characters}

We start by reviewing the proposal of \cite{Mathur:1988gt,Mukhi:1989qk} that the family of Feigin-Fuchs contour integrals, used in \cite{Dotsenko:1984nm,Dotsenko:1984ad} to describe sphere four-point functions of minimal models, can be put to an alternate use, namely computing the modular $S$-matrix. This proposal invokes the modular $\lambda$-function that maps (six copies of) the fundamental region of the torus onto the complex plane. Using this, it was argued that one can describe the characters of large classes of RCFT in terms of contour integrals. It was also noted that this representation permits explicit evaluation of the modular transformation matrices $S$ and $T$, which encode crucial information about the RCFT.

It has long been known \cite{Mathur:1988na, Naculich:1988xv, Mathur:1988gt} that the characters of 2-character RCFT can be expressed in terms of hypergeometric functions of $\lambda$. As an example, for a theory with two characters and Wronskian index $\ell=0$, we have:
\be
\begin{split}
\chi_0(\tau) &= \cN_0\, \left(\lambda(1-\lambda)\right)^{\frac{1}{6}-h} {}_2F_1\Big(\sfrac{1}{2}-h,\shalf-3h  ~\Big|~ 1-2h
~\Big|~\lambda\Big)\\[2mm]
\chi_1(\tau) &= \cN_1\, \left(\lambda(1-\lambda)\right)^{\frac{1}{6}+h} {}_2F_1\Big(\sfrac{1}{2}+h,\shalf+3h~\Big|~1+2h~\Big|~\lambda\Big)
\end{split}
\label{hyper}
\ee
where $h$ is the conformal dimension of the non-trivial primary, and the central charge is determined via the Riemann-Roch theorem to be $c=12h-2$. $\cN_0$ and $\cN_1$ are normalisations to be determined.
Finally, $\lambda$ is defined by:
\begin{equation}
\lambda (\tau) = \frac{\theta_2^4 (\tau)}{\theta_3^4 (\tau)} = 16\, q^{1/2} (1 - 8 q^{1/2} + 44 q + \dots)
\end{equation}
This is a series in half-integer powers of $q\equiv e^{2\pi i\tau}$. 
Now, the above hypergeometric functions have a contour integral representation:
\be
\begin{split}
F\Big(\sfrac{1}{2}-h,\shalf-3h  ~\Big|~ 1-2h
~\Big|~\lambda\Big)&=\frac{\Gamma(1-2h)}{\Gamma(\shalf-3h)\Gamma(\shalf +h)}
\int_1^\infty dt \Big[t(t-1)(t-\lambda)\Big]^{h-\half}\\
F\Big(\sfrac{1}{2}+h,\shalf+3h  ~\Big|~ 1+2h
~\Big|~\lambda\Big)&=
\big(\lambda(1-\lambda)\big)^{-2h}
\frac{\Gamma(1+2h)}{(\Gamma(\shalf+h))^2}
\int_0^\lambda dt \Big[t(1-t)(\lambda-t)\Big]^{h-\half}
\end{split}
\label{hypercontour}
\ee
This is defined by analytic continuation from $|\lambda|<1$ and from values of $h$ for which it converges. Note that for generic rational $h$ the integrand has branch points at $0,\lambda,1,\infty$ which need to be treated carefully. The factors in the integrand are ordered according to the integration range, so that there is no additional phase.

Inserting \eref{hypercontour} into \eref{hyper} and absorbing the $\Gamma$-function prefactors into new normalisations, we get:
\be
\begin{split}
\chi_0(\tau) &= N_0\, \left(\lambda(1-\lambda)\right)^{\frac{1}{6}-h} \int_1^\infty dt\, \Big[t(t-1)(t-\lambda)\Big]^{h-\half}\\[2mm]
\chi_1(\tau) &= N_1\, \left(\lambda(1-\lambda)\right)^{\frac{1}{6}-h} \int_0^{\lambda~} dt\, \Big[t(1-t)(\lambda-t)\Big]^{h-\half}
\end{split}
\label{charcontourtwo}
\ee
This is a simple and beautiful result, and provides useful information for us, as we will show below. Notice that the  prefactors and integrands are the same for both the characters, apart from normalisations. Thus the two characters differ only in the contour along which the integral is performed. 

Next, recall that the modular $\mathcal{T}$ and $\mathcal{S}$ transformations are the generators of the full modular group, under which the characters of a conformal field theory transform as follows:
\begin{equation}
\begin{split}
\mathcal{T} &: \chi_i(\tau) \to \chi_i(\tau+1)= e^{2 \pi i (- c/24 + h_i)} \chi_i(\tau)\\
\mathcal{S} &: \chi_i(\tau) \to \chi_i(-\sfrac{1}{\tau})=\sum_j \mathcal{S}_{ij} \chi_j(\tau)
\end{split}
\end{equation}
In terms of the modular $\lambda$-function, these transformations can be expressed as
\begin{equation}
\begin{split}
\mathcal{T} &: \lambda \to \frac{\lambda}{\lambda - 1}\\
\mathcal{S} &: \lambda \to 1 - \lambda
\end{split}
\label{TStrans}
\end{equation}
In particular the $\mathcal{S}$ transformation leaves the prefactor in \eref{charcontourtwo} invariant, but changes the integrand. By a simple change of variables this can be brought back to the original one, but with different contours of integration. The details are worked out in Appendix A. The result is an explicit formula that expresses each $\chi_i(-\frac{1}{\tau})$ in terms of linear combinations of the two original $\chi_i(\tau)$. Thus we have computed the modular transformation matrix $\mathcal{S}$. This is a huge improvement over the representation of characters as a $q$-series, where one can read off degeneracies easily but it is not possible to compute the modular $\mathcal{S}$-transformation properties.

In \cite{Mathur:1988gt,Mukhi:1989qk} a generalisation of \eref{charcontourtwo} was proposed based on inspiration from \cite{Dotsenko:1984nm,Dotsenko:1984ad} where integrals of this type were used in a different context. In this generalisation, the single integration variable $t$ is replaced by $n_1$ variables $t_i$ and $n_2$ variables $\tau_j$. For every pair of integers $A,A'$ where $0\le A\le n_1,~0\le A'\le n_2$, we define the multi-variable integration contours ${\cal C}_A$ and ${\cal C}_{A'}$ as follows:
\be
\begin{split}
\int_{{\cal C}_A}\prod_{i=1}^{n_1}dt_i &=
\int_1^\infty dt_{n_1}\cdots \int_1^{\infty}dt_{A+2}\int_1^{\infty} dt_{A+1}
\int_0^\lambda dt_{A} \cdots \int_0^{\lambda}dt_2\int_0^{\lambda} dt_1
\\
\int_{{\cal C}_{A'}}\prod_{j=1}^{n_2}d\tau_j &
=\int_1^\infty d\tau_{n_2} \cdots \int_1^{\infty}d\tau_{A'+2}\int_1^{\infty} d\tau_{A'+1}
\int_0^\lambda d\tau_{A'}\cdots \int_0^{\lambda}d\tau_{2}\int_0^{\lambda} d\tau_1
\end{split}
\ee
In order to avoid overlapping contours, we give each contour a distinct imaginary part, following the rule that:
\be
\Im(t_i)>\Im(t_k)~\hbox{for}~i>k,\qquad \Im(\tau_j)<\Im(\tau_l)~\hbox{for}~j>l
\ee 
In the manipulations below, we assume $\lambda$ is real and lies between 0 and 1. We may then extend it to be complex with $|\lambda|<1$, and finally analytically continue it to the complex plane.

Now, the proposal is that the following set of integrals are candidate characters of an RCFT:
\be
\begin{split}
J_{AA'}(a,b,\lambda)&\equiv
N_{AA'}\left(\lambda (1-\lambda)\right)^\alpha ~\int_{{\cal C}_A} \prod_{i=1}^{n_1}dt_i
\int_{{\cal C}_{A'}} \prod_{j=1}^{n_2}d\tau_j \\
&\quad\times \prod_{i=1}^{A} \Big[ t_i (1-t_i) (\lambda-t_i) \Big]^{a} 
\prod_{i=A+1}^{n_1} \Big[ t_i (t_i - 1) (t_i - \lambda) \Big]^{a} \\[2mm]
&\quad\times \prod_{j=1}^{A'} \Big[ \tau_j (1-\tau_j) (\lambda-\tau_j) \Big]^{b} 
\prod_{j=A'+1}^{n_2} \Big[ \tau_j (\tau_j-1) (\tau_j-\lambda) \Big]^{b} 
\\[2mm]
& \quad\times \prod_{1 \leq k < i \leq n_1} (t_i - t_k)^{-2a/b}
\prod_{1 \leq l < j \leq n_2} (\tau_j - \tau_l)^{-2b/a}
\prod_{i,j} (t_i - \tau_j)^{-2}
\end{split}
\label{calJdef}
\ee
where, as we will shortly see, $\alpha$ is a function of $a,b,n_1,n_2$ and the latter parameters determine the critical exponents of the theory.

Here the subscript $AA'$ can be thought of as a composite index taking $p=(n_1+1)(n_2+1)$ values. Thus there are $p$ functions in all.  To keep the notation simple we have not explicitly displayed the $n_1,n_2$ dependence of $J$. This can generally be read off from the context. The prefactors $N_{AA'}$ are normalisation factors that we will shortly discuss in more detail. It is also important to note that with the given ordering of terms in the integrand and the given integration contours, all brackets in the integrand are positive except the ones involving $(t_i-t_k)$ and $(\tau_j-\tau_l)$. The ordering for these brackets has been chosen so that the imaginary part is positive/negative for $t_i,\tau_j$ respectively.

For future use, we will often write:
\be
J_{AA'}(a,b,\lambda)=N_{AA'}\left(\lambda(1-\lambda)\right)^\alpha \hJ_{AA'}(a,b,\lambda)
\label{JJprime}
\ee
so that $\hJ_{AA'}$ denotes the ``pure'' integral without prefactors. A key feature of these integrals is that if we replace $\lambda$ by $1-\lambda$, which is the modular $S$-transformation, one has the result:
\be
\hJ_{AA'}(\lambda)=\sum_{B,B'} {\hat\mS}_{AA'\,BB'}\hJ_{BB'}(1-\lambda)
\ee
for some matrix $\hmS$. If we repeat the $\lambda\to 1-\lambda$ transformation we recover the original integrals, and therefore $\hmS^2=1$.

Now from \eref{JJprime} we see that the $\lambda$-dependent prefactor is invariant under this transformation. It follows that the normalised characters (written without hats) satisfy:
\be
J_{AA'}(\lambda)=\sum_{B,B'} \mS_{AA' \, BB'}J_{BB'}(1-\lambda)
\ee
where:
\be
\mS_{AA'\,BB'}=\frac{N_{AA'}}{N_{BB'}}\,{\hat\mS}_{AA'\,BB'}
\ee
This can be written as the matrix product:
\be
\mS= N\hmS N^{-1}
\label{StoShat}
\ee
where $N={\rm diag}(N_{AA'})$. From this it immediately follows that $\mS^2=1$ as well. 

Thus, the computation of the modular $\mS$-matrix has two parts: the calculation of $\hmS$ by manipulating the contour integrals, and the computation of the normalisations $N_{AA'}$. Each of these involves considerable work and we will carry them out separately.

As noted in \cite{Mukhi:1989qk}, the fact that this proposal actually works for known theories is quite surprising. There are two distinct surprises here. First of all, a generic $p$-character RCFT has discrete data consisting of the central charge $c$ and the $p-1$ conformal dimensions $h_i$.  For any given theory, these are rational points in a space of $p$ real dimensions. However the contour integrals give rise to functions having leading behaviours determined by two integer variables $n_1,n_2$ and two real dimensions labelled by $a$ and $b$. There is no a priori reason that any choice of $n_1,n_2,a,b$ in the contour will reproduce the discrete data for an $p$-character theory.

To see this in more detail, we should compute the $\cal T$ transformation of the $J_{AA'}$ to extract the critical exponents. This is straightforward to do by inserting \eref{TStrans} into the integral representation. One finds that the integrals are eigenfunctions of $\cal T$ with eigenvalues that are phases giving the critical exponents. A simpler way to achieve the same thing is to compute the leading power of $\lambda$ in each of the $J_{AA'}$. This has been done in \cite{Mukhi:1989qk} with the result that:
\be
J_{AA'}\sim \lambda^{\alpha+\Delta_{AA'}}
\ee
where:
\be
\begin{split}
\alpha &= \frac13\Big(-n_1(1+3a) -n_2(1+3b)+\frac{a}{b}n_1(n_1-1)+ \frac{b}{a}n_2(n_2-1)+2n_1n_2\Big) \\
\Delta_{AA'} &= A(1+2a)+A'(1+2b)-\frac{a}{b} A(A-1)-\frac{b}{a}A'(A'-1)-2AA'
\end{split}
\label{Deltadef}
\ee
Using the fact that $\lambda\sim \sqrt{q}$ for small $\lambda$, we see that:
\be
J_{AA'}\sim q^{\half(\alpha+\Delta_{AA'})}
\label{critexp}
\ee
Whenever $J_{AA'}$ are the characters of an RCFT, the above expression gives us the critical exponents of the theory. Thus we have the identification:
\be
\half (\alpha+\Delta_{AA'})=-\frac{c}{24}+h_i
\ee
where $c$ is the central charge and $h_i, i=0,1,2,\cdots p-1$ are the conformal dimensions associated to the characters, with $h_0=0$ corresponding to the identity character. The total number of characters $p$ is identified with $(n_1+1)(n_2+1)$. Note in particular that for the identity character we have $A=A'=0$ and hence $\Delta_{AA'}=0$, leading to the relations:
\be
c=-12\alpha,\quad h_i=\half \Delta_{AA'}
\label{calpha}
\ee 
where the composite label $AA'$ takes the same $p$ values as the index $i$. 

Now the Riemann-Roch theorem tells us that:
\be
\sum_{i=0}^{n-1}\left(-\frac{c}{24}+h_i\right)=\frac{n(n-1)}{12}-\frac{\ell}{6}
\ee
where $\ell$ is the Wronskian index of the CFT. From \eref{Deltadef} we find that for all theories whose characters are of the form $J_{AB}$, the Wronskian index is $\ell=0$. 

The surprise arises when we consider known theories with $\ell=0$. There are infinitely many such theories \cite{Mathur:1988gt} and one can check whether their critical exponents fit the formula \eref{critexp}. This is assured for theories with two or three characters, as already noted in \cite{Mathur:1988gt}, but from four characters onwards, the system appears overdetermined. However, in \cite{Mukhi:1989qk} it was verified that the critical exponents can indeed be fitted with \eref{critexp} for the following infinite series:

\bi

\item
SU(2)$_k$ WZW models for all $k$.

\item
All $c<1$ minimal models (unitary as well as non-unitary). These are parametrised by two integers $p$ and $p'$ such that:
\be
c=1-\frac{6(p-p')^2}{pp'}
\ee

\item
SU(N)$_1$ WZW models for all N.

\item
The $D$ and $E$ series of (non-diagonal) invariants for SU(2)$_k$ as well as $c<1$ minimal models. 

\item
The SU(3)$_2$ WZW model. 

\ei

As an example, the critical exponents of the $(p,p')$ minimal model are reproduced by choosing $n_1=p-2,n_2=\half(p'-3)$ and $a=\frac{3p'}{4p}-1, b=\frac{2p}{p'}-\frac{3}{2}$ where we have chosen $p'$ odd without loss of generality. This is quite striking and suggests the contour integral representation is somehow ``tailor-made'' to describe the minimal models and several other infinite series of theories. 

That was the first surprise. Now we turn to the second one. It is known \cite{Mathur:1988na, Mathur:1988gt} that for any theory with $\ell=0$ and $\le 5$ characters, the critical exponents completely determine the characters via a modular linear differential equation (MLDE). This ensures  that for every case with $<6$ characters where we have successfully reproduced the critical exponents from contour integrals, the characters are correctly described by the contour integrals. However for theories with $\ge 6$ characters there is no such proof. Rather, some evidence for this was provided in \cite{Mukhi:1989qk} by explicitly computing one coefficient in the $q$-series for the identity character. For SU(2)$_k$ this coefficient should be equal to 3 for all $k$, while for any minimal model it should be 0 (because $L_{-1}$ on the vacuum is a null vector), and for SU(N)$_1$ it should be N$^2-1$. These three predictions were verified in \cite{Mukhi:1989qk}. It follows that the six-character theories SU(2)$_5$, the minimal models with $(p,p')=(4,5), (2,13), (3,7)$, and the SU(10)$_1$, SU(11)$_1$ models are all described by the contour-integral representation. For the remaining SU(2)$_k$, the minimal models with more than six characters and the SU(N)$_1$ models with N$>6$, this calculation provides one piece of evidence in support of the conjecture. 

Finally, in \cite{Mukhi:1989qk}, use was made of certain elements of the modular transformation matrix $\mS$ computed in \cite{Dotsenko:1984nm, Dotsenko:1984ad} using the contour integral method for the following cases:

\bi
\item
$\mS_{jj'}$ with $j'=\frac{k}{2}$ for SU(2)$_k$.

\item
$\mS_{rr',ss'}$ with $r=s=p-1,r'=s'=p'-2$ for the $p,p'$ minimal models. 
\ei
These agree with the known results for these cases. 

In the present work we take the proposal of \cite{Mathur:1988gt,Mukhi:1989qk} further in several ways. First, we systematise the computation of the modular $\mS$-matrix. This is done in two stages. The first is to calculate the monodromy matrix $\hmS$ by manipulating contours, and the other is to compute the normalisations by evaluating specific integrals. We carry out  each of these steps in turn. It may be noted that only the first of these stages was relevant for \cite{Dotsenko:1984nm, Dotsenko:1984ad}, because their work deals with correlators rather than characters. 
Moreover this stage was carried out step by step, one contour at a time, which quickly becomes very tedious. As a result, the authors of these references presented the monodromy matrix only for $(n_1,n_2)=(2,0)$ and gave results for certain entries/rows in the general case. We re-visit this method (restricting to $n_2=0$ in the present paper) and develop an elegant algorithmic procedure which expresses the result as a sum over paths. This is a significant simplification -- for any finite value the entire $\mS$-matrix can be easily computed by hand or by writing a short programme. We illustrate this by computing $\mS$ for the cases $n_1=2,3$ (three- and four-character cases) and could easily to do more except that the $\mS$-matrices quickly become too large to print.

On the way we examine to what extent the matrix $\mS$ can be used as a classification tool. We start with the 2-character case, for which $n_1=1,n_2=0$. We examine for which values of the parameters $a,\rho$ the fusion rules, determined from the $S$-matrix via Verlinde's formula, are non-negative integers. This actually leads us to re-discover known CFT's as well as the quasi-characters proposed in \cite{Chandra:2018pjq}. From 3 characters onwards one again finds constraints on the parameters $a, \rho$, but the formulae unfortunately appear intractable. 

Second, we apply our results to compute degeneracies and multiplicities for the case of 2,3 and 4 characters. This is particularly interesting in the context of quasi-characters, as well as for exotic RCFT's like the coset theories defined in \cite{Gaberdiel:2016zke}. 

Third, we consider specific cases beyond 5 characters to provide additional evidence for the conjectures of \cite{Mukhi:1989qk}. Our algorithm is able to easily compute generic entries of the $\mS$-matrix for quite large numbers of characters. We illustrate this by calculating the full $\mS$-matrix for the 8-character theory SU(2)$_7$. We show that this matrix numerically agrees with that derived from the Kac-Weyl formula for these characters. This leads us to think that for general SU(2)$_k$ theories it may be possible to sum the paths explicitly and reproduce the entire Kac-Weyl formula, which would be very convincing evidence that the contour integrals provide a correct description of RCFT characters.

Let us establish some conventions for the class of examples with $n_1$ arbitrary but $n_2=0$, which will be the main focus of many of our calculations below. In this case the only dependence of the integrands on $b$ is through the power $-\frac{a}{b}$ which can be treated as an independent variable in addition to $a$. For these cases we will denote $-\frac{a}{b}$ by $\rho$ and treat $a,\rho$ as the parameters. The integrals $J_{A0}$ will for brevity be denoted $J_A$. Also we will use the following notation for trigonometric functions in the rest of the paper: $\mf s(x)\equiv \sin(\pi x),\mf c(x)\equiv \cos(\pi x)$.

\section{Computation of the modular $\mS$-matrix}

\subsection{Monodromy of the integrals}

As we mentioned at the outset, a key feature of the contour integral representation is that it permits computation of the modular $\mS$-matrix via contour deformation. Here we systematise this computation and arrive at a description as a sum over paths, which renders the computation very explicit and allows us to go beyond the results of \cite{Mathur:1988gt,Mukhi:1989qk} which were taken in turn from \cite{Dotsenko:1984nm, Dotsenko:1984ad}. We will, however, limit us to the sub-class of contour integrals with just one type of integration variable ($t_i$), or in other words with $n_2=0$ in \eref{calJdef}. Renaming $n_1$ as $n$, the corresponding integrals have the form:
\be
\begin{split}
  \hat{J}_A(\lambda) &= \int_1^\infty dt_n \int_1^\infty dt_{n-1} \cdots \int_0^\lambda dt_A \cdots \int_0^\lambda dt_1 \nn \\
  &\quad\quad\times\prod_{i=1}^A \big[t_i (1-t_i)(\lambda -t_i)\big]^a \prod_{i=A+1}^n \big[t_i (t_i-1)(t_i-\lambda)\big]^a \prod_{0\le k<i\le n}(t_i-t_k)^{2\rho}
  \label{oneparamJdef}
\end{split}
\ee
Notice that $\hat{J}_A$ has $A$ contours running from 0 to $\lambda$ and the remaining $n-A$ from 1 to $\infty$. Thus there are altogether $p=n+1$ contour integrals, which are going to be candidates to describe a $p$-character CFT. A generic integral in this set can be represented by the diagram:
\begin{align}
    \begin{tikzpicture}
     \node at (-6,-0.03) {$\hat{J}_A(\lambda)=$};
     \draw (-5,0)--(5,0);
     \filldraw (-3,0) circle (2pt) node[anchor=north]{0};
     \filldraw (0,0) circle (2pt) node[anchor=north]{$\lambda$};
     \filldraw (3,0) circle (2pt)  node[anchor=north]{1};
     \draw[->] (-3,0).. controls (-1.5, 0.2) .. (-0.1,0.05);
     \draw[->] (-3,0).. controls (-1.5, 1.2) .. (-0.05,0.1);
     \draw[dotted] (-1.5,0.2)--(-1.5,0.93);
     \node at (-1.35,0.45) {$A$};
     \draw[->] (3,0).. controls (3.2, 1) .. (5,1);
     \draw[->] (3,0).. controls (3.2, 0.2) .. (5,0.2);
     \draw[dotted] (4,0.2)--(4,1);
     \node at (4.6,0.6) {$n-A$};
    \end{tikzpicture}
\end{align}

One now takes each of these contours and deforms it to its complement along the real axis. This is valid because the contribution from a semi-circular contour that lies entirely in the upper half-plane or lower half-plane is zero. Thus the deformation can be carried out in two ways, once in the upper half-plane and once in the lower half-plane. Now the integrand is branched at points where any of the $t_i$ is equal to $0,\lambda,1,\infty$ and also when any pair $t_i,t_k$ is equal. Thus the deformed contour requires specification of a phase in each segment, and these phases differ in the upper and lower half-planes. The last step is to take the difference of these two deformed integrals (each of which is equal to the original one) with factors that cancel out the contribution from the segment 0 to $\lambda$. For those unfamiliar with this ``unfolding'' technique, a simple example is provided in Appendix \ref{unfolding}.

The result of the above procedure is that the original contour has disappeared and is replaced by two contours, one in each of the segments $(1,\lambda)$ and $(0,-\infty)$. Now if we carry out the procedure $s$ times, sequentially on the $A$ contours in the $(0,\lambda)$ region, we will be left with only $A-s$ of these contours. The remaining $s$ will be distributed in the regions $(0,-\infty)$ and $(1,\lambda)$. We may introduce an integer $m$ such that $\frac{s-m}{2}$ contours are in $(0,-\infty)$ and $\frac{s+m}{2}$ are in $(1,\lambda)$. Clearly $m$ takes the values $-s,-s+2,\cdots,s-2,s$. The contour configuration for a fixed $(s,m)$ will be labelled $\msc V_{s,m}$ and, as long as $s < A$, it looks like the following:
\begin{align}
    &\begin{tikzpicture}
     \node at (-6,-0.03) {$\msc V_{s,m}=$};
     \draw (-5,0)--(5,0);
     \filldraw (-3,0) circle (2pt) node[anchor=north]{0};
     \filldraw (0,0) circle (2pt) node[anchor=north]{$\lambda$};
     \filldraw (3,0) circle (2pt)  node[anchor=north]{1};
     \draw[->] (-3,0).. controls (-1.5, 0.2) .. (-0.1,0.05);
     \draw[->] (-3,0).. controls (-1.5, 1.2) .. (-0.05,0.1);
     \draw[dotted] (-1.5,0.2)--(-1.5,0.95);
     \node at (-1.05,0.45) {$A-s$};
     \draw[->] (3,0).. controls (3.2, 1) .. (5,1);
     \draw[->] (3,0).. controls (3.2, 0.2) .. (5,0.2);
     \draw[dotted] (4,0.2)--(4,1);
     \node at (4.6,0.6) {$n-A$};
     \draw[->] (-3,0).. controls (-3.2, 1) .. (-5,1);
     \draw[->] (-3,0).. controls (-3.2, 0.2) .. (-5,0.2);
     \draw[dotted] (-4,0.2)--(-4,1);
     \node at (-4.6,0.6) {$\frac{s-m}{2}$};
     \draw[<-] (0.1,0.05).. controls (1.5, 0.2) .. (3,0);
     \draw[<-] (0.05,0.1).. controls (1.5, 1.2) .. (3,0);
     \draw[dotted] (1.5,0.2)--(1.5,0.95);
     \node at (1.81,0.5) {$\frac{s+m}{2}$};
     \end{tikzpicture}
\end{align}

Once a particular $\msc V_{s,m}$ has been reached (with $s<A$), a further unfolding of a single contour in the $(\lambda,0)$ region leads to either $\msc V_{s+1,m-1}$ or $\msc V_{s+1,m+1}$ depending on whether the unfolded contour is replaced by another contour in the $(0,-\infty)$  region or the $(\lambda,1)$ region respectively. We now seek a recursive relation between $V_{s,m}$ and its successor diagrams  $\msc V_{s+1,m-1}$ and $\msc V_{s+1,m+1}$. This follows from the following diagrammatic steps. First we deform the contour in the UHP, to get:
\begin{align}
  &\begin{tikzpicture}
     \node at (-6,-0.03) {$\msc V_{s,m}=$};
     \draw (-5,0)--(5,0);
     \filldraw (-3,0) circle (2pt) node[anchor=north]{0};
     \filldraw (0,0) circle (2pt) node[anchor=north]{$\lambda$};
     \filldraw (3,0) circle (2pt)  node[anchor=north]{1};
     \draw[->] (-3,0).. controls (-1.5, 0.2) .. (-0.1,0.05);
     \draw[->] (-3,0).. controls (-1.5, 1.2) .. (-0.05,0.1);
     \draw[dotted] (-1.5,0.2)--(-1.5,0.95);
     \node at (-1.35,0.45) {$A-s$};
     \draw[->] (3,0).. controls (3.2, 1) .. (5,1);
     \draw[->] (3,0).. controls (3.2, 0.2) .. (5,0.2);
     \draw[dotted] (4,0.2)--(4,1);
     \node at (4.6,0.6) {$n-A$};
     \draw[->] (-3,0).. controls (-3.2, 1) .. (-5,1);
     \draw[->] (-3,0).. controls (-3.2, 0.2) .. (-5,0.2);
     \draw[dotted] (-4,0.2)--(-4,1);
     \node at (-4.6,0.6) {$\frac{s-m}{2}$};
     \draw[<-] (0.1,0.05).. controls (1.5, 0.2) .. (3,0);
     \draw[<-] (0.05,0.1).. controls (1.5, 1.2) .. (3,0);
     \draw[dotted] (1.5,0.2)--(1.5,0.95);
     \node at (1.81,0.5) {$\frac{s+m}{2}$};
   \end{tikzpicture} \nn \\
  &\begin{tikzpicture}
     \node at (-7,-0.03) {$= e^{i\pi(a+(A-s-1)2\rho)}$};
     \draw (-5,0)--(5,0);
     \filldraw (-3,0) circle (2pt) node[anchor=north]{0};
     \filldraw (0,0) circle (2pt) node[anchor=north]{$\lambda$};
     \filldraw (3,0) circle (2pt)  node[anchor=north]{1};
     \draw[->] (-3,0).. controls (-1.5, 0.2) .. (-0.1,0.05);
     \draw[->] (-3,0).. controls (-1.5, 1.2) .. (-0.05,0.1);
     \draw[dotted] (-1.5,0.2)--(-1.5,0.95);
     \node at (-1.5,0.45) {$A-s-1$};
     \draw[->] (3,0).. controls (3.2, 1) .. (5,1);
     \draw[->] (3,0).. controls (3.2, 0.2) .. (5,0.2);
     \draw[dotted] (4,0.2)--(4,1);
     \node at (4.6,0.6) {$n-A$};
     \draw[->] (-3,0).. controls (-3.2, 1) .. (-5,1);
     \draw[->] (-3,0).. controls (-3.2, 0.2) .. (-5,0.2);
     \draw[->] (-3,0).. controls (-3.2, 1.4) .. (-5,1.4);
     \draw[dotted] (-4,0.2)--(-4,1);
     \node at (-4.6,0.6) {$\frac{s-m}{2}$};
     \draw[<-] (0.1,0.05).. controls (1.5, 0.2) .. (3,0);
     \draw[<-] (0.05,0.1).. controls (1.5, 1.2) .. (3,0);
     \draw[dotted] (1.5,0.2)--(1.5,0.95);
     \node at (1.81,0.5) {$\frac{s+m}{2}$};
   \end{tikzpicture} \nn \\  
  &\begin{tikzpicture}
     \node at (-7,-0.03) {$+ e^{-i\pi(a+(s+m)\rho)}$};
     \draw (-5,0)--(5,0);
     \filldraw (-3,0) circle (2pt) node[anchor=north]{0};
     \filldraw (0,0) circle (2pt) node[anchor=north]{$\lambda$};
     \filldraw (3,0) circle (2pt)  node[anchor=north]{1};
     \draw[->] (-3,0).. controls (-1.5, 0.2) .. (-0.1,0.05);
     \draw[->] (-3,0).. controls (-1.5, 1.2) .. (-0.05,0.1);
     \draw[dotted] (-1.5,0.2)--(-1.5,0.95);
     \node at (-1.5,0.45) {$A-s-1$};
     \draw[->] (3,0).. controls (3.2, 1) .. (5,1);
     \draw[->] (3,0).. controls (3.2, 0.2) .. (5,0.2);
     \draw[dotted] (4,0.2)--(4,1);
     \node at (4.6,0.6) {$n-A$};
     \draw[->] (-3,0).. controls (-3.2, 1) .. (-5,1);
     \draw[->] (-3,0).. controls (-3.2, 0.2) .. (-5,0.2);
     \draw[dotted] (-4,0.2)--(-4,1);
     \node at (-4.6,0.6) {$\frac{s-m}{2}$};
     \draw[<-] (0.1,0.05).. controls (1.5, 0.2) .. (3,0);
     \draw[<-] (0.05,0.1).. controls (1.5, 1.2) .. (3,0);
     \draw[<-] (0.05,0.2).. controls (1.5, 1.6) .. (3,0);
     \draw[dotted] (1.5,0.2)--(1.5,0.95);
     \node at (1.81,0.5) {$\frac{s+m}{2}$};
   \end{tikzpicture} \nn \\
    &\begin{tikzpicture}
     \node at (-7,-0.03) {$+ e^{-i\pi(2a+(s+m+2n-2A)\rho)}$};
     \draw (-5,0)--(5,0);
     \filldraw (-3,0) circle (2pt) node[anchor=north]{0};
     \filldraw (0,0) circle (2pt) node[anchor=north]{$\lambda$};
     \filldraw (3,0) circle (2pt)  node[anchor=north]{1};
     \draw[->] (-3,0).. controls (-1.5, 0.2) .. (-0.1,0.05);
     \draw[->] (-3,0).. controls (-1.5, 1.2) .. (-0.05,0.1);
     \draw[dotted] (-1.5,0.2)--(-1.5,0.95);
     \node at (-1.5,0.45) {$A-s-1$};
     \draw[->] (3,0).. controls (3.2, 1) .. (5,1);
     \draw[->] (3,0).. controls (3.2, 0.2) .. (5,0.2);
     \draw[<-] (3,0.1).. controls (3.2, 1.4) .. (5,1.4);
     \draw[dotted] (4,0.2)--(4,1);
     \node at (4.6,0.6) {$n-A$};
     \draw[->] (-3,0).. controls (-3.2, 1) .. (-5,1);
     \draw[->] (-3,0).. controls (-3.2, 0.2) .. (-5,0.2);
     \draw[dotted] (-4,0.2)--(-4,1);
     \node at (-4.6,0.6) {$\frac{s-m}{2}$};
     \draw[<-] (0.1,0.05).. controls (1.5, 0.2) .. (3,0);
     \draw[<-] (0.05,0.1).. controls (1.5, 1.2) .. (3,0);
     \draw[dotted] (1.5,0.2)--(1.5,0.95);
     \node at (1.81,0.5) {$\frac{s+m}{2}$};
   \end{tikzpicture} \label{eq:up}
\end{align}
Next, we repeat the procedure in the LHP to find:
\begin{align}
  &\begin{tikzpicture}
     \node at (-6,-0.03) {$\msc V_{s,m}=$};
     \draw (-5,0)--(5,0);
     \filldraw (-3,0) circle (2pt) node[anchor=north]{0};
     \filldraw (0,0) circle (2pt) node[anchor=north]{$\lambda$};
     \filldraw (3,0) circle (2pt)  node[anchor=north]{1};
     \draw[->] (-3,0).. controls (-1.5, 0.2) .. (-0.1,0.05);
     \draw[->] (-3,0).. controls (-1.5, 1.2) .. (-0.05,0.1);
     \draw[dotted] (-1.5,0.2)--(-1.5,0.95);
     \node at (-1.35,0.45) {$A-s$};
     \draw[->] (3,0).. controls (3.2, 1) .. (5,1);
     \draw[->] (3,0).. controls (3.2, 0.2) .. (5,0.2);
     \draw[dotted] (4,0.2)--(4,1);
     \node at (4.6,0.6) {$n-A$};
     \draw[->] (-3,0).. controls (-3.2, 1) .. (-5,1);
     \draw[->] (-3,0).. controls (-3.2, 0.2) .. (-5,0.2);
     \draw[dotted] (-4,0.2)--(-4,1);
     \node at (-4.6,0.6) {$\frac{s-m}{2}$};
     \draw[<-] (0.1,0.05).. controls (1.5, 0.2) .. (3,0);
     \draw[<-] (0.05,0.1).. controls (1.5, 1.2) .. (3,0);
     \draw[dotted] (1.5,0.2)--(1.5,0.95);
     \node at (1.81,0.5) {$\frac{s+m}{2}$};
   \end{tikzpicture} \nn \\
  &\begin{tikzpicture}
     \node at (-6.7,-0.03) {$= e^{-i\pi(a+(s-m)\rho)}$};
     \draw (-5,0)--(5,0);
     \filldraw (-3,0) circle (2pt) node[anchor=north]{0};
     \filldraw (0,0) circle (2pt) node[anchor=north]{$\lambda$};
     \filldraw (3,0) circle (2pt)  node[anchor=north]{1};
     \draw[->] (-3,0).. controls (-1.5, 0.2) .. (-0.1,0.05);
     \draw[->] (-3,0).. controls (-1.5, 1.2) .. (-0.05,0.1);
     \draw[dotted] (-1.5,0.2)--(-1.5,0.95);
     \node at (-1.5,0.45) {$A-s-1$};
     \draw[->] (3,0).. controls (3.2, 1) .. (5,1);
     \draw[->] (3,0).. controls (3.2, 0.2) .. (5,0.2);
     \draw[dotted] (4,0.2)--(4,1);
     \node at (4.6,0.6) {$n-A$};
     \draw[->] (-3,0).. controls (-3.2, 1) .. (-5,1);
     \draw[->] (-3,0).. controls (-3.2, 0.2) .. (-5,0.2);
     \draw[->] (-3,0).. controls (-3.2, -0.4) .. (-5,-0.4);
     \draw[dotted] (-4,0.2)--(-4,1);
     \node at (-4.6,0.6) {$\frac{s-m}{2}$};
     \draw[<-] (0.1,0.05).. controls (1.5, 0.2) .. (3,0);
     \draw[<-] (0.05,0.1).. controls (1.5, 1.2) .. (3,0);
     \draw[dotted] (1.5,0.2)--(1.5,0.95);
     \node at (1.81,0.5) {$\frac{s+m}{2}$};
   \end{tikzpicture} \nn \\  
  &\begin{tikzpicture}
     \node at (-6.7,-0.03) {$+ e^{i\pi(a+(A-s-1)2\rho)}$};
     \draw (-5,0)--(5,0);
     \filldraw (-3,0) circle (2pt) node[anchor=north]{0};
     \filldraw (0,0) circle (2pt) node[anchor=north]{$\lambda$};
     \filldraw (3,0) circle (2pt)  node[anchor=north]{1};
     \draw[->] (-3,0).. controls (-1.5, 0.2) .. (-0.1,0.05);
     \draw[->] (-3,0).. controls (-1.5, 1.2) .. (-0.05,0.1);
     \draw[dotted] (-1.5,0.2)--(-1.5,0.95);
     \node at (-1.5,0.45) {$A-s-1$};
     \draw[->] (3,0).. controls (3.2, 1) .. (5,1);
     \draw[->] (3,0).. controls (3.2, 0.2) .. (5,0.2);
     \draw[dotted] (4,0.2)--(4,1);
     \node at (4.6,0.6) {$n-A$};
     \draw[->] (-3,0).. controls (-3.2, 1) .. (-5,1);
     \draw[->] (-3,0).. controls (-3.2, 0.2) .. (-5,0.2);
     \draw[dotted] (-4,0.2)--(-4,1);
     \node at (-4.6,0.6) {$\frac{s-m}{2}$};
     \draw[<-] (0.1,0.05).. controls (1.5, 0.2) .. (3,0);
     \draw[<-] (0.05,0.1).. controls (1.5, 1.2) .. (3,0);
     \draw[<-] (0.1,-0.05).. controls (1.5, -0.5) .. (3,0);
     \draw[dotted] (1.5,0.2)--(1.5,0.95);
     \node at (1.81,0.5) {$\frac{s+m}{2}$};
   \end{tikzpicture} \nn \\
  &\begin{tikzpicture}
     \node at (-7,-0.03) {$+e^{i\pi(2a+(2A-s+m-2)\rho)}$};
     \draw (-5,0)--(5,0);
     \filldraw (-3,0) circle (2pt) node[anchor=north]{0};
     \filldraw (0,0) circle (2pt) node[anchor=north]{$\lambda$};
     \filldraw (3,0) circle (2pt)  node[anchor=north]{1};
     \draw[->] (-3,0).. controls (-1.5, 0.2) .. (-0.1,0.05);
     \draw[->] (-3,0).. controls (-1.5, 1.2) .. (-0.05,0.1);
     \draw[dotted] (-1.5,0.2)--(-1.5,0.95);
     \node at (-1.5,0.45) {$A-s-1$};
     \draw[->] (3,0).. controls (3.2, 1) .. (5,1);
     \draw[->] (3,0).. controls (3.2, 0.2) .. (5,0.2);
     \draw[<-] (3.1,-0.1).. controls (3.2, -0.5) .. (5,-0.5);
     \draw[dotted] (4,0.2)--(4,1);
     \node at (4.6,0.6) {$n-A$};
     \draw[->] (-3,0).. controls (-3.2, 1) .. (-5,1);
     \draw[->] (-3,0).. controls (-3.2, 0.2) .. (-5,0.2);
     \draw[dotted] (-4,0.2)--(-4,1);
     \node at (-4.6,0.6) {$\frac{s-m}{2}$};
     \draw[<-] (0.1,0.05).. controls (1.5, 0.2) .. (3,0);
     \draw[<-] (0.05,0.1).. controls (1.5, 1.2) .. (3,0);
     \draw[dotted] (1.5,0.2)--(1.5,0.95);
     \node at (1.81,0.5) {$\frac{s+m}{2}$};
   \end{tikzpicture} \label{eq:down}
\end{align}
To remove the $\infty\to1$ piece of the contour, we take the linear combination:
\be
e^{i\pi (2a + (n+m-1)\rho)}\times \eqref{eq:up} - e^{-i\pi (2a + (n+m-1)\rho)}\times\eqref{eq:down}
\ee
Thus we can express the diagram $\msc V_{s,m}$ in terms of the two successor diagrams $\msc V_{s+1,m\pm1}$.
\be
\begin{split}
  \msc V_{s,m} &= \frac{e^{i\pi(2a + (n+m-1)\rho)}e^{i\pi(a+(A-s-1)2\rho)}-e^{-i\pi(2a + (n+m-1)\rho)}e^{-i\pi(a+(s-m)\rho)}}{2i\;\mf s(2a+(n+m-1)\rho)}\msc V_{s+1,m-1}\\
  &\quad + \frac{e^{i\pi(2a + (n+m-1)\rho)}e^{-i\pi(a+(s+m)\rho)}-e^{-i\pi(2a + (n+m-1)\rho)}e^{i\pi(a+(A-s-1)2\rho)}}{2i\;\mf s(2a+(n+m-1)\rho)}\msc V_{s+1,m+1} \\
  &=\frac{e^{i\pi\left(A-\frac{3s-m}{2}-1\right)\rho}\;\mf s(3a+(n+A-\left(\frac{s-m}{2}\right)-2)\rho)}{\mf s(2a + (n+m-1)\rho)}\msc V_{s+1,m-1}\\
  &\quad + \frac{e^{i\pi\left(A-\frac{3s+m}{2}-1\right)\rho}\;\mf s(a+(n-A+\left(\frac{s+m}{2}\right))\rho)}{\mf s(2a + (n+m-1)\rho)}\msc V_{s+1,m+1}  
\end{split}
\ee
where we recall that $\mf s(x)\equiv \sin(\pi x)$. 

The above formula is true only for $s<A$. After $A$ steps there will be no $0\to\lambda$ contours left. Thereafter we must unfold the remaining $n-A$ contours in the $(1,\infty)$ region by deforming each contour in two ways and removing the contributions from the region $(\lambda,0)$. Let us find the effect of carrying out one unfolding starting from a value of $s\ge A$. At this stage the system looks like this:
\begin{align}
\begin{tikzpicture}
     \node at (-6,-0.03) {$\msc V_{s,m} = $};
     \draw (-5,0)--(5,0);
     \filldraw (-3,0) circle (2pt) node[anchor=north]{0};
     \filldraw (0,0) circle (2pt) node[anchor=north]{$\lambda$};
     \filldraw (3,0) circle (2pt)  node[anchor=north]{1};
     \draw[->] (3,0).. controls (3.2, 1) .. (5,1);
     \draw[->] (3,0).. controls (3.2, 0.2) .. (5,0.2);
     \draw[dotted] (4,0.2)--(4,1);
     \node at (4.6,0.6) {$n-s$};
     \draw[->] (-3,0).. controls (-3.2, 1) .. (-5,1);
     \draw[->] (-3,0).. controls (-3.2, 0.2) .. (-5,0.2);
     \draw[dotted] (-4,0.2)--(-4,1);
     \node at (-4.6,0.6) {$\frac{s-m}{2}$};
     \draw[<-] (0.1,0.05).. controls (1.5, 0.2) .. (3,0);
     \draw[<-] (0.05,0.1).. controls (1.5, 1.2) .. (3,0);
     \draw[dotted] (1.5,0.2)--(1.5,0.95);
     \node at (1.81,0.5) {$\frac{s+m}{2}$};
   \end{tikzpicture} 
\end{align}
Unfolding one of the contours in the $(1,\infty)$ region we find:
\begin{align}
  &\begin{tikzpicture}
     \node at (-7.3,-0.03) {$\msc V_{s,m} = e^{i\pi(3a+(2n-s+m-2)\rho)}$};
     \draw (-5,0)--(5,0);
     \filldraw (-3,0) circle (2pt) node[anchor=north]{0};
     \filldraw (0,0) circle (2pt) node[anchor=north]{$\lambda$};
     \filldraw (3,0) circle (2pt)  node[anchor=north]{1};
     \draw[->] (3,0).. controls (3.2, 1) .. (5,1);
     \draw[->] (3,0).. controls (3.2, 0.2) .. (5,0.2);
     \draw[dotted] (4,0.2)--(4,1);
     \node at (4.8,0.6) {$n-s-1$};
     \draw[->] (-3,0).. controls (-3.2, 1) .. (-5,1);
     \draw[->] (-3,0).. controls (-3.2, 0.2) .. (-5,0.2);
     \draw[->] (-3,0).. controls (-3.2, 1.4) .. (-5,1.4);
     \draw[dotted] (-4,0.2)--(-4,1);
     \node at (-4.6,0.6) {$\frac{s-m}{2}$};
     \draw[<-] (0.1,0.05).. controls (1.5, 0.2) .. (3,0);
     \draw[<-] (0.05,0.1).. controls (1.5, 1.2) .. (3,0);
     \draw[dotted] (1.5,0.2)--(1.5,0.95);
     \node at (1.81,0.5) {$\frac{s+m}{2}$};
   \end{tikzpicture} \nn \\  
  &\begin{tikzpicture}
     \node at (-7,-0.03) {$+ e^{i\pi(2a+(2n-s+m-2)\rho)}$};
     \draw (-5,0)--(5,0);
     \filldraw (-3,0) circle (2pt) node[anchor=north]{0};
     \filldraw (0,0) circle (2pt) node[anchor=north]{$\lambda$};
     \filldraw (3,0) circle (2pt)  node[anchor=north]{1};
     \draw[<-] (-2.95,0.05).. controls (-1.5, 1) .. (0,0);
     \draw[->] (3,0).. controls (3.2, 1) .. (5,1);
     \draw[->] (3,0).. controls (3.2, 0.2) .. (5,0.2);
     \draw[dotted] (4,0.2)--(4,1);
     \node at (4.8,0.6) {$n-s-1$};
     \draw[->] (-3,0).. controls (-3.2, 1) .. (-5,1);
     \draw[->] (-3,0).. controls (-3.2, 0.2) .. (-5,0.2);
     \draw[dotted] (-4,0.2)--(-4,1);
     \node at (-4.6,0.6) {$\frac{s-m}{2}$};
     \draw[<-] (0.1,0.05).. controls (1.5, 0.2) .. (3,0);
     \draw[<-] (0.05,0.1).. controls (1.5, 1.2) .. (3,0);
     \draw[dotted] (1.5,0.2)--(1.5,0.95);
     \node at (1.81,0.5) {$\frac{s+m}{2}$};
   \end{tikzpicture} \nn \\
    &\begin{tikzpicture}
     \node at (-7,-0.03) {$+ e^{i\pi(a+(n-s-1)2\rho)}$};
     \draw (-5,0)--(5,0);
     \filldraw (-3,0) circle (2pt) node[anchor=north]{0};
     \filldraw (0,0) circle (2pt) node[anchor=north]{$\lambda$};
     \filldraw (3,0) circle (2pt)  node[anchor=north]{1};
     \draw[->] (3,0).. controls (3.2, 1) .. (5,1);
     \draw[->] (3,0).. controls (3.2, 0.2) .. (5,0.2);
     \draw[dotted] (4,0.2)--(4,1);
     \node at (4.8,0.6) {$n-s-1$};
     \draw[->] (-3,0).. controls (-3.2, 1) .. (-5,1);
     \draw[->] (-3,0).. controls (-3.2, 0.2) .. (-5,0.2);
     \draw[dotted] (-4,0.2)--(-4,1);
     \node at (-4.6,0.6) {$\frac{s-m}{2}$};
     \draw[<-] (0.1,0.05).. controls (1.5, 0.2) .. (3,0);
     \draw[<-] (0.05,0.1).. controls (1.5, 1.2) .. (3,0);
     \draw[<-] (0.05,0.2).. controls (1.5, 1.6) .. (3,0);
     \draw[dotted] (1.5,0.2)--(1.5,0.95);
     \node at (1.81,0.5) {$\frac{s+m}{2}$};
   \end{tikzpicture} \label{eq:GeqA_up}
\end{align}
and 
\begin{align}
  &\begin{tikzpicture}
     \node at (-7,-0.03) {$\msc V_{s.m} = e^{-i\pi(3a+2s\rho)}$};
     \draw (-5,0)--(5,0);
     \filldraw (-3,0) circle (2pt) node[anchor=north]{0};
     \filldraw (0,0) circle (2pt) node[anchor=north]{$\lambda$};
     \filldraw (3,0) circle (2pt)  node[anchor=north]{1};
     \draw[->] (3,0).. controls (3.2, 1) .. (5,1);
     \draw[->] (3,0).. controls (3.2, 0.2) .. (5,0.2);
     \draw[dotted] (4,0.2)--(4,1);
     \node at (4.6,0.6) {$n-s-1$};
     \draw[->] (-3,0).. controls (-3.2, 1) .. (-5,1);
     \draw[->] (-3,0).. controls (-3.2, 0.2) .. (-5,0.2);
     \draw[->] (-3,0).. controls (-3.2, -0.4) .. (-5,-0.4);
     \draw[dotted] (-4,0.2)--(-4,1);
     \node at (-4.6,0.6) {$\frac{s-m}{2}$};
     \draw[<-] (0.1,0.05).. controls (1.5, 0.2) .. (3,0);
     \draw[<-] (0.05,0.1).. controls (1.5, 1.2) .. (3,0);
     \draw[dotted] (1.5,0.2)--(1.5,0.95);
     \node at (1.81,0.5) {$\frac{s+m}{2}$};
   \end{tikzpicture} \nn \\  
  &\begin{tikzpicture}
     \node at (-6.7,-0.03) {$+\;e^{-i\pi(2a+(s+m)\rho)}$};
     \draw (-5,0)--(5,0);
     \filldraw (-3,0) circle (2pt) node[anchor=north]{0};
     \filldraw (0,0) circle (2pt) node[anchor=north]{$\lambda$};
     \filldraw (3,0) circle (2pt)  node[anchor=north]{1};
     \draw[<-] (-2.95,-0.05).. controls (-1.5, -0.5) .. (0,0);
     \draw[->] (3,0).. controls (3.2, 1) .. (5,1);
     \draw[->] (3,0).. controls (3.2, 0.2) .. (5,0.2);
     \draw[dotted] (4,0.2)--(4,1);
     \node at (4.8,0.6) {$n-s-1$};
     \draw[->] (-3,0).. controls (-3.2, 1) .. (-5,1);
     \draw[->] (-3,0).. controls (-3.2, 0.2) .. (-5,0.2);
     \draw[dotted] (-4,0.2)--(-4,1);
     \node at (-4.6,0.6) {$\frac{s-m}{2}$};
     \draw[<-] (0.1,0.05).. controls (1.5, 0.2) .. (3,0);
     \draw[<-] (0.05,0.1).. controls (1.5, 1.2) .. (3,0);
     \draw[dotted] (1.5,0.2)--(1.5,0.95);
     \node at (1.81,0.5) {$\frac{s+m}{2}$};
   \end{tikzpicture}  \nn \\
  &\begin{tikzpicture}
     \node at (-6.7,-0.03) {$+\;e^{-i\pi(a+(s+m)\rho)}$};
     \draw (-5,0)--(5,0);
     \filldraw (-3,0) circle (2pt) node[anchor=north]{0};
     \filldraw (0,0) circle (2pt) node[anchor=north]{$\lambda$};
     \filldraw (3,0) circle (2pt)  node[anchor=north]{1};
     \draw[->] (3,0).. controls (3.2, 1) .. (5,1);
     \draw[->] (3,0).. controls (3.2, 0.2) .. (5,0.2);
     \draw[dotted] (4,0.2)--(4,1);
     \node at (4.8,0.6) {$n-s-1$};
     \draw[->] (-3,0).. controls (-3.2, 1) .. (-5,1);
     \draw[->] (-3,0).. controls (-3.2, 0.2) .. (-5,0.2);
     \draw[dotted] (-4,0.2)--(-4,1);
     \node at (-4.6,0.6) {$\frac{s-m}{2}$};
     \draw[<-] (0.1,0.05).. controls (1.5, 0.2) .. (3,0);
     \draw[<-] (0.05,0.1).. controls (1.5, 1.2) .. (3,0);
     \draw[<-] (0.1,-0.05).. controls (1.5, -0.5) .. (3,0);
     \draw[dotted] (1.5,0.2)--(1.5,0.95);
     \node at (1.81,0.5) {$\frac{s+m}{2}$};
   \end{tikzpicture} \label{eq:GeqA_down}
\end{align}
Taking the linear combination 
\be
e^{i\pi (2a + (n+m-1)\rho)}\times\eqref{eq:GeqA_down} - e^{-i\pi (2a + (n+m-1)\rho)}\times
\eqref{eq:GeqA_up}
\ee
 we end up with:
\be
\begin{split}
  \msc V_{s,m} &= \frac{e^{i\pi(2a + (n+m-1)\rho)}e^{-i\pi(3a+2s\rho)}-e^{-i\pi(2a + (n+m-1)\rho)}e^{i\pi(3a+(2n-s+m-2)\rho)}}{2i\;\mf s(2a+(n+m-1)\rho)}\msc V_{s+1,m-1} \\
  &\,\, + \frac{e^{i\pi(2a + (n+m-1)\rho)}e^{-i\pi(a+(s+m)\rho)}-e^{-i\pi(2a + (n+m-1)\rho)}e^{i\pi(a+(n-s-1)2\rho)}}{2i\;\mf s(2a+(n+m-1)\rho)}\msc V_{s+1,m+1} \\
  &=-\frac{e^{i\pi\left(n-\frac{3s-m}{2}-1\right)\rho}\;\mf s(a+\left(\frac{s-m}{2}\right)\rho)}{\mf s(2a + (n+m-1)\rho)}\msc V_{s+1,m-1} + \frac{e^{i\pi\left(n-\frac{3s+m}{2}-1\right)\rho}\;\mf s(a+\left(\frac{s+m}{2}\right)\rho)}{\mf s(2a + (n+m-1)\rho)}\msc V_{s+1,m+1}
\end{split}
\ee
which is valid for $s\ge A$.

Thus we are led to define the following coefficients:
\be
\begin{split}
s<A:\quad
L^-_{s,m} &= \frac{e^{i\pi\left(A-\frac{3s-m}{2}-1\right)\rho}\;\mf s(3a+(n+A-\left(\frac{s-m}{2}\right)-2)\rho)}{\mf s(2a + (n+m-1)\rho)}\\
  L^+_{s,m} &= \frac{e^{i\pi\left(A-\frac{3s+m}{2}-1\right)\rho}\;\mf s(a+(n-A+\left(\frac{s+m}{2}\right))\rho)}{\mf s(2a + (n+m-1)\rho)}\\
s\ge A:\quad 
L^-_{s,m} &= -\frac{e^{i\pi\left(n-\frac{3s-m}{2}-1\right)\rho}\;\mf s(a+\left(\frac{s-m}{2}\right)\rho)}{\mf s(2a + (n+m-1)\rho)}\\
  L^+_{s,m} &= \frac{e^{i\pi\left(n-\frac{3s+m}{2}-1\right)\rho}\;\mf s(a+\left(\frac{s+m}{2}\right)\rho)}{\mf s(2a + (n+m-1)\rho)}
  \end{split}
  \ee
In terms of these, we have the recursion relation:
\be
\msc V_{s,m}=L^-_{s,m}\msc V_{s+1,m-1}+L^+_{s,m}\msc V_{s+1,m+1}
\ee
Therefore one unfolding can be represented as the following diagram
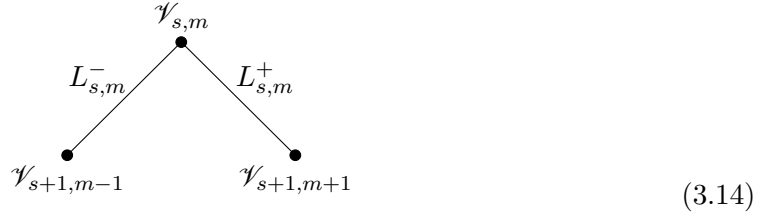
\begin{align}
  \begin{tikzpicture}
    \filldraw (0,0) circle (2pt) node[anchor=south]{$\msc V_{s,m}$};
    \filldraw (-1.5,-1.5) circle (2pt) node[anchor=north]{$\msc V_{s+1,m-1}$};
    \filldraw (1.5,-1.5) circle (2pt) node[anchor=north]{$\msc V_{s+1,m+1}$};
    \draw (0,0)--(-1.5,-1.5);
    \draw (0,0)--(1.5,-1.5);
     \node at (-1.1,-0.5) {$L^-_{s,m}$};
    \node at (1.1,-0.5) {$L^+_{s,m}$};
  \end{tikzpicture}
\end{align}
where both the links are directed downwards.
Starting from $\hat J_A (\lambda)$ which can be identified as $\msc V_{0,0}$, $n$-unfoldings can be represented as the following graph:
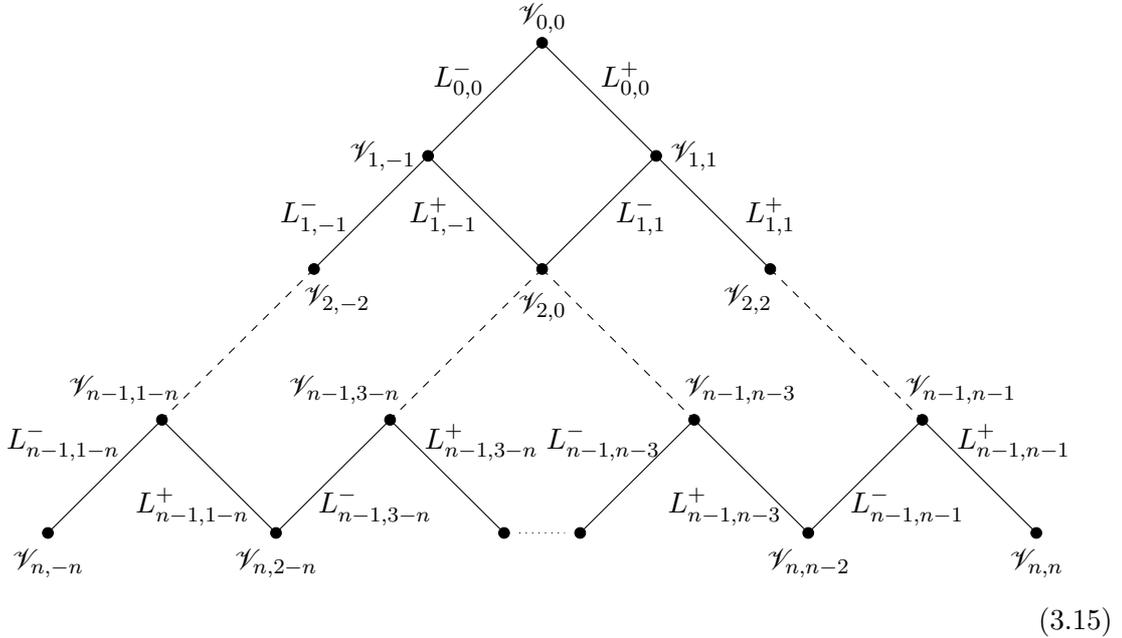
\be
  \begin{split}
    &\begin{tikzpicture}
       \filldraw (0,0) circle (2pt) node[anchor=south]{$\msc V_{0,0}$};
       \filldraw (-1.5,-1.5) circle (2pt) node[anchor=east]{};
       \node at (-2.1,-1.5) {$\msc V_{1,-1}$};
       \filldraw (1.5,-1.5) circle (2pt) node[anchor=west]{};
       \node at (2,-1.5) {$\msc V_{1,1}$};
       \filldraw (-3,-3) circle (2pt) node[anchor=east]{};
       \node at (-2.7,-3.4) {$\msc V_{2,-2}$};
       \filldraw (0,-3) circle (2pt) node[anchor=south]{};
       \node at (0,-3.5) {$\msc V_{2,0}$};
       \filldraw (3,-3) circle (2pt) node[anchor=west]{};
       \node at (2.7,-3.4) {$\msc V_{2,2}$};
       \filldraw (-5,-5) circle (2pt) node[anchor=west]{};
       \node at (-5.5,-4.6) {$\msc V_{n-1,1-n}$};
       \filldraw (-2,-5) circle (2pt) node[anchor=east]{};
       \node at (-2.6,-4.6) {$\msc V_{n-1,3-n}$};
       \filldraw (2,-5) circle (2pt) node[anchor=south]{};
       \node at (2.6,-4.6) {$\msc V_{n-1,n-3}$};
       \filldraw (5,-5) circle (2pt) node[anchor=west]{};
       \node at (5.5,-4.6) {$\msc V_{n-1,n-1}$};
       \filldraw (-6.5,-6.5) circle (2pt) node[anchor=west]{};
       \node at (-6.5,-6.9) {$\msc V_{n,-n}$};
       \filldraw (-3.5,-6.5) circle (2pt) node[anchor=east]{};
       \node at (-3.5,-6.9) {$\msc V_{n,2-n}$};
       \filldraw (3.5,-6.5) circle (2pt) node[anchor=south]{};
       \node at (3.5,-6.9) {$\msc V_{n,n-2}$};
       \filldraw (6.5,-6.5) circle (2pt) node[anchor=west]{};
       \node at (6.5,-6.9) {$\msc V_{n,n}$};
       \filldraw (-0.5,-6.5) circle (2pt) node[anchor=west]{};
       \filldraw (0.5,-6.5) circle (2pt) node[anchor=east]{};
       \draw (0,0)--(-1.5,-1.5);
       \draw (0,0)--(1.5,-1.5);
       \draw (-1.5,-1.5)--(-3,-3);
       \draw (-1.5,-1.5)--(0,-3);
       \draw (1.5,-1.5)--(0,-3);
       \draw (1.5,-1.5)--(3,-3);
       \draw[dashed] (-3,-3)--(-5,-5);
       \draw[dashed] (0,-3)--(-2,-5);
       \draw[dashed] (0,-3)--(2,-5);
       \draw[dashed] (3,-3)--(5,-5);
       \draw (-5,-5)--(-6.5,-6.5);
       \draw (-2,-5)--(-3.5,-6.5);
       \draw (2,-5)--(3.5,-6.5);
       \draw (5,-5)--(6.5,-6.5);
       \draw (-5,-5)--(-3.5,-6.5);
       \draw (-2,-5)--(-0.5,-6.5);
       \draw (2,-5)--(0.5,-6.5);
       \draw (5,-5)--(3.5,-6.5);
       \draw[dotted] (-0.3,-6.5)--(0.3,-6.5);
       \node at (-1.1,-0.5) {$L^-_{0,0}$};
       \node at (1.1,-0.5) {$L^+_{0,0}$};
       \node at (-3,-2.3) {$L^-_{1,-1}$};
       \node at (-1.3,-2.3) {$L^+_{1,-1}$};
       \node at (1.3,-2.3) {$L^-_{1,1}$};    
       \node at (3,-2.3) {$L^+_{1,1}$};
       \node at (-6.3,-5.3) {$L^-_{n-1,1-n}$};
       \node at (-4.6,-6.15) {$L^+_{n-1,1-n}$};
       \node at (-2.2,-6.15) {$L^-_{n-1,3-n}$};    
       \node at (-0.8,-5.3) {$L^+_{n-1,3-n}$};
       \node at (0.8,-5.3) {$L^-_{n-1,n-3}$};
       \node at (2.4,-6.15) {$L^+_{n-1,n-3}$};
       \node at (4.8,-6.15) {$L^-_{n-1,n-1}$};
       \node at (6.2,-5.3) {$L^+_{n-1,n-1}$};
     \end{tikzpicture}
    \label{coolgraph}
  \end{split}
\ee
Notice that the vertices in the final row are the integrals $\hat J_A (1-\lambda)$ such that
\be
  \msc V_{n,2A-n} = (-1)^n\hat J_A (1-\lambda)
\ee
where the sign on the RHS comes from the change of variables $t_i\to 1-t_i$ (see Appendix \ref{unfolding} for a simple example). 
The algorithm to compute the $\hat{\mc S}$ matrix is as follows:
\begin{enumerate}
  \item For the row in the $\hat {\mc S}$-matrix corresponding to $\hat{J}_A$, identify $\msc V_{0,0}$ with $\hat{J}_A (\lambda)$.
  \item To compute the element $\hat{\mc S}_{A \, B}$, trace a path in the graph \eqref{coolgraph} starting from $\msc V_{0,0}$ to $\msc V_{n,2B-n}$ while multiplying the contribution from each link on the path.
  \item Sum over the contributions from all such paths to obtain $\hat{\mc S}_{A \, B}$.

\end{enumerate}

Now that we have shown how to compute $\hmS$ in a simple algorithmic fashion, we go on to calculate the normalisations of the integrals $\hJ_A$. 

%% ============= Reached here =============

%% \begin{align}
%%   L^+_{s,m} &=
%%   \begin{cases}
%%     \frac{ e^{i \pi \left( A-\frac{3s+m}{2}-1 \right)} \mf s \left( a + \left( n-A+\frac{s+m}{2} \right) \rho \right) }{ \mf s \left( 2a + (n+m-1) \rho \right) }, & \hspace*{0.825cm} s < A \\[0.5cm]
%%     \frac{ e^{i \pi \left( n-\frac{3s+m}{2}-1 \right)} \mf s \left( a + \left( \frac{s+m}{2} \right) \rho \right) }{ \mf s \left( 2a + (n+m-1) \rho \right) }, & \hspace*{0.825cm} s \geq A
%%   \end{cases} \nn \\[0.3cm]
%%   L^-_{s,m} &=
%%   \begin{cases}
%%     \frac{ e^{i \pi \left( A-\frac{3s-m}{2}-1 \right)} \mf s \left( 3a + \left( n+A-\frac{s-m}{2}-2 \right) \rho \right) }{ \mf s \left( 2a + (n+m-1) \rho \right) }, & s < A \\[0.5cm]
%%     -\frac{ e^{i \pi \left( n-\frac{3s-m}{2}-1 \right)} \mf s \left( a + \left( \frac{s-m}{2} \right) \rho \right) }{ \mf s \left( 2a + (n+m-1) \rho \right) }, & s \geq A
%%   \end{cases}
%% \end{align}

%% ========================= This part is over ========================= 

\subsection{Normalisation of the integrals}

The normalisations $N_A$ were originally introduced in 
\eref{calJdef}. Recall that the second index of $N_{AA'}$ has been dropped because we are working with the class of integrals having $n_2=0$. Thus, \eref{StoShat} becomes:
\be
\mS_{AB}=\frac{N_A}{N_B}\,\hmS_{AB}
\label{StoShat.2}
\ee
Thus, once we compute the normalisations we will have finally determined the modular $\mS$-matrix.

The factors $N_A$ are defined by requiring that the coefficient of the lowest power of $q$ is the ground-state degeneracy $D_A$ of the corresponding character, which is a non-negative integer such that:
\be
\chi_A(q)= D_A~ q^{\half(\alpha+\Delta_A)}(1+\cO(q))
\ee
Note that for a genuine RCFT, we should have $D_0=1$ so that the identity character is non-degenerate (this can be relaxed for theories of IVOA type \cite{Kawasetsu:2014} and for quasi-characters \cite{Chandra:2018pjq}).

To calculate the $N_A$, we only need the leading behaviour in $q$, corresponding to taking the $\lambda \to 0$ limit. To compute this, it will be convenient to first define the ordered integral:
\be
\begin{split}
  I_A(\lambda) &= N_A \left(\lambda(1-\lambda)\right)^\alpha \int_1^\infty dt_n \int_1^{t_n} dt_{n-1} \cdots \int_1^{t_{A+2}} dt_{A+1} \int_0^\lambda dt_A \int_0^{t_A} dt_{A-1} \cdots \int_0^{t_2} dt_1 \\
  &\quad\quad\times\prod_{i=1}^A \big[t_i (1-t_i)(\lambda -t_i)\big]^a \prod_{i=A+1}^n \big[t_i (t_i-1)(t_i-\lambda)\big]^a \prod_{0\le k<i\le n}(t_i-t_k)^{2\rho}
  \label{oneparamIdef}
\end{split}
\ee
The integrand and pre-factors are identical to those in $J_A$, and the only difference is that the integration range for $I_A$ has $t_1<t_2<\cdots <t_n$. The relation between $J_A (\lambda)$ to $I_A (\lambda)$ as given in \cite{Dotsenko:1984nm} (and re-derived here in Appendix \ref{JtoI_App} for convenience) is:
\be
\begin{split}
  J_A (\lambda) &= \left(\prod_{k=1}^{A-1} e^{i\pi k\rho}\, \frac{\mf s((k+1)\rho)}{\mf s(\rho)} \prod_{l=1}^{n-A-1} e^{i\pi l \rho}\,\frac{\mf s((l+1)\rho)}{\mf s(\rho)} \right)I_A(\lambda) \\
  &\equiv \Theta_A(\rho)\, I_A(\lambda)
\end{split}
\label{JtoI}
\ee
For $A=0,1$ the product over $k$ is absent, and similarly for $A=n-1, n$ the product over $l$ is absent. 

In view of the ordered integrals, the factor $(t_i-t_k)$ is always non-negative. Thus we can place a modulus sign around it if we like.
Thereafter we are free to extend the integral to be un-ordered.
This will simply count the integration region $A!$ and $(n-A)!$ times.
Hence we have:
\be
\begin{split}
I_A(\lambda)
=&\frac{N_A}{A!(n-A)!}\left(\lambda(1-\lambda)\right)^\alpha
\int_1^\infty\!\! dt_{n} \cdots\int_1^\infty dt_{A+1}
\int_0^\lambda dt_A \cdots\int_0^\lambda \!\! dt_1  \\[2mm]
&\times \prod_{i=1}^A \big[t_i (1-t_i)(\lambda -t_i)\big]^a \prod_{i=A+1}^n \big[t_i (t_i-1)(t_i-\lambda)\big]^a \prod_{i<k}\big|t_i-t_k\big|^{2\rho}
\end{split}
\ee
Next, we make the substitutions $t_i=\lambda u_i$ for $i=1,2,\cdots A$, and $t_i= \frac{1}{u_i}$ for $i=A+1,A+2,\cdots,n$. The integral then becomes:
\be
\begin{split}
I_A(\lambda)
=&\frac{N_A}{A!(n-A)!}\,\lambda^{\alpha+A\big(1+2a+\rho(A-1)\big)} (1-\lambda)^\alpha
\int_0^1 du_n\cdots \int_0^1 du_1\\[2mm]
&\times \prod_{i=1}^{A} 
\Big[u_i(1-u_i)(1-\lambda u_i)\Big]^a 
\prod_{i=A+1}^{n} 
u_i^{-2-3a-2\rho(n-1)} \Big[(1-u_i)(1-\lambda u_i)\Big]^a  \\[2mm]
&\times \prod_{1\le i<k\le A}\big|u_i-u_k\big|^{2\rho}
\prod_{A+1\le i<k\le n}\big|u_i-u_k\big|^{2\rho}
\prod_{\substack{1\le i\le A,\\[1.5mm]A+1\le k\le n}}\big|1-\lambda u_i u_k\big|^{2\rho}
\end{split}
\ee
Notice that the leading power of $\lambda$, which has been factored out of the integral, is just $\Delta_{A0}$ as defined in \eref{Deltadef}. Henceforth we denote this by $\Delta_A$.  

In the above form it is easy to compute the leading behaviour as $\lambda \to 0$, and one finds:
\be
  \begin{split}
    I_A(\lambda \to 0)
    &=\frac{N_A}{A!(n-A)!}\,\lambda^{\alpha+\Delta_A} \int_0^1 du_A\cdots \int_0^1 du_1
     ~ \prod_{i=1}^{A} 
    \Big[u_i(1-u_i)\Big]^a
     \prod_{1\le i<k\le A}\big|u_i-u_k\big|^{2\rho}\\[2mm]
    &\quad \times \int_0^1 du_n\cdots \int_0^1 du_{A+1}
    \prod_{i=A+1}^{n} 
    u_i^{-2-3a-2\rho(n-1)} (1-u_i)^a
    \prod_{A+1\le i<k\le n}\big|u_i-u_k\big|^{2\rho}\\[2mm]
    &=
    \frac{N_A}{A!(n-A)!}\,(16\, \sqrt{q}\,)^{\alpha+\Delta_A}~ S_A \big( a+1,a+1,\rho \big)\, S_{n-A} \big( -1-3a-2\rho(n-1),a+1,\rho \big)
  \end{split}
\ee
In this limit the integral has factorised into two parts, and we have identified each of these as a Selberg integral. This integral is defined as:
\be
  \begin{split}
    S_m (\alpha,\beta,\gamma) &= \int_0^1 du_m \cdots \int_0^1 du_1~ \prod_{i=1}^m u_i^{\alpha-1} (1-u_i)^{\beta-1} \prod_{1 \leq i < j \leq m} \big|u_i-u_j\big|^{2\gamma}\\[2mm]
    &= m!~\prod_{k=1}^m \frac{\Gamma(k\gamma)}{\Gamma(\gamma)} ~\prod_{k=0}^{m-1} \frac{\Gamma(\alpha+k\gamma)\, \Gamma(\beta+k\gamma)}{\Gamma(\alpha+\beta+(m+k-1)\gamma)}
  \end{split}
\label{Selberg}
\ee
Demanding this to be equal to the ground-state degeneracy, $D_A$ of the corresponding character we fix the normalisation to be
\be
  N_A = \frac{16^{-(\alpha+\Delta_A)}\, A! (n-A)! \, D_A}{\Theta_A(\rho)\,S_A \big( a+1,a+1,\rho \big) \; S_{n-A} \big( -1-3a-2\rho(n-1),a+1,\rho \big)}
\label{normcomp}
\ee

\subsection{Final result}

One can now insert the results from the previous two sections, namely \eref{coolgraph} and \eref{normcomp}, into \eref{StoShat.2} to obtain the modular $\mc S$ matrix for any desired case. Note that it will contain the undetermined factor $D_A$. This can be found in multiple ways: (i) if one knows the chiral algebra of the theory then the $D_A$ would be known (our method can still provide a useful way to compute $\mS$), (ii) in the spirit of \cite{Mathur:1988na, Naculich:1988xv,Mathur:1988gt} one can guess the $D_A$ by requiring an integral $q$-series, (iii) following \cite{Gaberdiel:2016zke,Hampapura:2016mmz} one can determine the $D_A$ for coset theories via the bilinear coset relation.

\section{Applications: Theories with $\le 4$ characters}

As explained at the outset, all $p$-character RCFT's with $p\le 3$ and $\ell=0$ must have contour-integral representations for their characters. For theories with $p=4$ (also $p=5$), there will be a contour-integral description as long as the critical exponents can be identified using \eref{Deltadef}, specialised to the case we are considering in this paper ($n_1=n,n_2=0$), where they become:
\be
\begin{split}
\alpha &= -n\left(\frac{(n-1)}{3}\rho +a+\frac13\right)\\
\Delta_{A} &= A\big(1+2a+ \rho (A-1)\big)
\label{Deltadef.2}
\end{split}
\ee
In view of \eref{calpha} it also follows that such theories will have a central charge and conformal dimensions:
\be
c=4n\big((n-1)\rho +3a+1\big),\quad h_A=\half A\big(1+2a+ \rho (A-1)\big)
\label{cform}
\ee
We now examine by turns these two classes of theories ($p\le 3$ and $p=4$) in the light of the formulae derived in the present paper.

Before proceeding further, it is important to highlight another subtle point. As emphasised in \cite{Mathur:1988na} and subsequent works, the number of characters is not necessarily equal to the number of primary fields under the (extended) chiral algebra. When more than one primary has the same character, we say that the character comes with a ``multiplicity'' $M_A$. Simple examples with $M_A > 1$ arise when there are complex primaries ($M_A=2$, since the field and its complex conjugate have the same character), and in theories with a special symmetry like SO(8) triality ($M_A=3$, where three different representations have the same character). However there are also more complicated situations where large values of $M_A$ arise \cite{Bae:2017kcl}. 

Whenever there are multiplicities $M_A>1$, the matrix $\mS$ computed from the characters is not unitary but instead satisfies:
\begin{equation}
  \mathcal{S}^\dagger 
  \begin{pmatrix} 1~ & 0 & 0 & \cdots & 0\\ 
  0~ & M_1 &0 & \cdots & 0\\
  0~ & 0 & M_2 & \cdots & 0 \\
  \vdots & \vdots & \vdots & \ddots & \vdots \\
  0~ & 0 & 0 & \cdots & M_n
    \end{pmatrix} 
        \mathcal{S} = 
          \begin{pmatrix} 1~ & 0 & 0 & \cdots & 0\\ 
  0~ & M_1 &0 & \cdots & 0\\
  0~ & 0 & M_2 & \cdots & 0 \\
  \vdots & \vdots & \vdots & \ddots & \vdots \\
  0~ & 0 & 0 & \cdots & M_n
    \end{pmatrix}
  \label{SMatProp}
\end{equation}

In general it can be rather difficult to identify the $M_A$ without prior knowledge of a corresponding CFT. Moreover, whenever $M_A>1$ we cannot use the Verlinde formula directly on $\mS$ to determine the fusion rules. Instead, as in \cite{Mathur:1988gt}, we must first ``expand'' $\mS$ to a unitary matrix whose size is the same as the number of primary fields (not characters), and then apply the Verlinde formula. 

If one assumes $M_A=1$ for all $A$, thereby classifying only those theories with this property (i.e. theories with $n$ characters and exactly $n$ primary fields) then of course $\mS$ is unitary. In this case one may require integrality of the fusion rules to obtain constraints on the allowed values of $a,\rho$. We will work this out in some special cases. 

In addition to characters of RCFT's, the contour integral approach is useful in understanding quasi-characters, introduced in \cite{Chandra:2018pjq}. These are solutions of the MLDE whose coefficients are integral though not necessarily positive. Thus they are not candidate characters for an actual conformal field theory. Nevertheless, at least for the case of two characters (the only case studied in \cite{Chandra:2018pjq}), they are useful as building blocks of candidate CFT characters, which they generate via linear combinations. In what follows, we will in particular revisit and complete the discussion of \cite{Chandra:2018pjq} for the two-character case.

We now embark on a systematic survey of the contour integral representation and its modular $\mS$-matrix for the cases $p=2,3,4$.

\subsection{2 characters}

We start with the simplest application of our procedure, namely the case $n=1$. While this is essentially worked out in previous papers \cite{Mathur:1988gt, Naculich:1988xv,Chandra:2018pjq} we present it here in a more complete and systematic form. In this case \eref{cform} gives:
\be
c=12a+4,\quad h=a+\half
\label{chtwo}
\ee
The normalisation factors of \eref{normcomp} become:
\be
\begin{split}
N_0& = \frac{(16)^{a+\frac13} D_0}{\beta(a+1,-3a-1)}\\
N_1& = \frac{(16)^{-a-\frac23}D_1}{\beta(a+1,a+1)}
\label{norms.2}
\end{split}
\ee
and:
\be
\hmS =\begin{pmatrix} \frac{\mf s(a)}{\mf s(2a)} & -\frac{\mf s(a)} {\mf s(2a)} \\[2mm]
  -\frac{\mf s(3a)}{\mf s(2a)} & -\frac{\mf s(a)}{\mf s(2a)}
  \end{pmatrix}
\ee
Notice that $\hmS$ squares to 1, as expected. Now using \eref{StoShat.2} we find:
\be
\mS=  \begin{pmatrix} \frac{\mf s(a)}{\mf s(2a)} & -\frac{\mf s(a)}{\mf s(2a)} \frac{N_0}{N_1} \vspace{0.2cm} \\
    -\frac{\mf s(3a)}{\mf s(2a)} \frac{N_1}{N_0} & -\frac{\mf s(a)}{\mf s(2a)}
  \end{pmatrix}
\ee
This matrix agrees with the two-character modular $\mathcal{S}$ matrix derived in \cite{Mathur:1988gt,Naculich:1988xv} derived using hypergeometric functions. 

Now we implement the requirement that:
\begin{equation}
  \mathcal{S}^\dagger \begin{pmatrix} 1 & 0 \\ 0 & M \end{pmatrix} \mathcal{S} = \begin{pmatrix} 1 & 0 \\ 0 & M \end{pmatrix}
  \label{SMatProp.2}
\end{equation}
where we have denoted the multiplicity $M_1$ of the non-identity character by $M$ for simplicity.
This gives:
\be
  \frac{N_1}{N_0} =  \pm\frac{1}{\sqrt{M}}~\sqrt{\frac{\mf s(a)}{\mf s(3a)}}
\label{unitar}
  \ee
The sign ambiguity above has to be resolved case by case depending on the value of $a$, which (from \eref{norms.2}) tells us the correct sign of the ratio 
$\frac{N_1}{N_0}$.

Anticipating the relevance of the above not just to RCFT's but also to quasi-characters, let us recall the results of \cite{Chandra:2018pjq} in this context. The second order MLDE with $\ell=0$ has solutions whose critical exponents $c,h$ fall into four distinct classes. It is sufficient to list the values of $a$ for these classes since this determines $c,h$ via \eref{chtwo}. The four classes are:
\be
\begin{split}
\text{Lee-Yang: }\quad a &= \frac{r}{5}-\frac{3}{10},\quad r\ne 4\text{ mod }5\\[2mm]
A_1\!:\quad a &= \frac{r}{2}-\frac{1}{4}\\[2mm]
A_2\!:\quad a &= \frac{r}{3}-\frac{1}{6},\quad r\ne 2\text{ mod }3\\[2mm]
D_4\!:\quad a &= r
\end{split}
\label{fclasses}
\ee

Next we assume $M=1$ and apply the Verlinde formula:
\be
  \mc N_{ijk} = \sum_l \frac{\mc S_{il} \mc S_{jl} \mc S^*_{kl}}{\mc S_{0l}}
\ee
Unitarity of the $\mS$-matrix, which we have already implemented above, ensures that $\cN_{0jk}=\delta_{jk}$. The only remaining fusion rule coefficient is:
\begin{equation}
\begin{split}
\mathcal{N}_{111} &= \frac{\mathcal{S}_{10}^3}{\mathcal{S}_{00}} + \frac{\mathcal{S}_{11}^3}{\mathcal{S}_{01}}\\
&= \frac{N_1}{N_0}\frac{\mf s(a)^2}{\mf s(2a)^2}\left[1-\left(\frac{N_1}{N_0}\right)^2 \left(\frac{\mf s(3a)}{\mf s(a)}\right)^3 \right]\\
&= \pm\frac{\mf s(a)^\frac52}{\mf s(3a)^\half \mf s(2a)^2}\left[1- \left(\frac{\mf s(3a)}{\mf s(a)}\right)^2 \right]
\end{split}
\end{equation}
where in the last line we have eliminated $\frac{N_1}{N_0}$ from \eref{unitar}.

Let us start by checking for theories that satisfy $\cN_{111}=0$. This implies:
\be
\frac{\mf s(3a)}{\mf s(a)}=\pm 1
\ee
which in turn gives:
\be
a=\frac{r}{2}-\frac14
\ee
for any integer $r$. From \eref{cform}, it follows that their central charges will be:
\be
c=6r+1
\ee
For $r=0,1$ these are the known WZW models of $A_{1,1}$ and $E_{7,1}$ \cite{Mathur:1988na} while for all other cases we have re-discovered the quasi-characters in the $A_1$ fusion class \cite{Chandra:2018pjq}.
Thus we have re-discovered an infinite family of quasi-characters just by assuming a specific consistent set of fusion rules. 

The next logical possibility is $\cN_{111}=\pm 1$. In this case one finds the set of values:
\be
a=\frac{r}{5}-\frac{3}{10}, \quad r \ne 4\text{ mod } 5
\ee
Via \eref{calpha} it follows that:
\be
c=\frac25(6r+1), \quad r \ne 4\text{ mod } 5
\ee
Here we have re-discovered the Lee-Yang series of quasi-characters, of which the first few elements correspond to actual CFT: for $n=0$ one has the Lee-Yang theory (in the ``unitary presentation'' where the lower dimension is identified with the identity character), for $r=1,2$ one finds the $G_{2,1}$ and $F_{4,1}$ WZW models and for $r=3$ we get the famous IVOA of central charge $\frac{38}{5}$. It is easy to verify that the sign of $N_{111}$ is $-,+,+,-$ for these four cases respectively, as already noted in \cite{Mathur:1988gt}.

We did not find the remaining quasi-character series ($A_2$ and $D_4$ classes) because they correspond to the case of multiplicity $M>1$. In this case the Verlinde formula cannot be applied with a $2\times 2$ matrix $\mS$. We also did not find any rational solution for $N_{111}=\pm 2$ etc, which is in any case ruled out by the analysis of \cite{Christe:1988xy}. From our point of view the intriguing fact is that just from the modular $\mS$-matrix and the Verlinde formula we are able to reproduce all quasi-characters for the cases without multiplicity, without ever solving any MLDE and invoking integrality of the coefficients in the solution.

We now compute the degeneracy ratio $\frac{D_1}{D_0}$ for all the quasi-character series. This was determined empirically in specific cases in \cite{Chandra:2018pjq} but no general formula was written down there. For this purpose we will no longer need to use the Verlinde formula, so we can re-introduce the multiplicity $M$ and consider all four series of quasi-characters.  Combining \eref{norms.2} and \eref{unitar}, we have:
\be
\frac{D_1}{D_0}=\frac{16^{2a+1}}{\sqrt{M}}\abs{\frac{\beta(a+1,a+1)}{\beta(a+1,-3a-1)}\sqrt\frac{\mf s(a)}{\mf s(3a)}}
\ee
The results for the four series of quasi-characters are then:
\be
\begin{split}
  \text{Lee-Yang series: }\qquad
  \frac{D_1}{D_0}&=16^{\frac{3}{10}(r+1)}
  \left(\sfrac{\sqrt5+1}{\sqrt5-1}\right)^{\pm\half}
    \abs{\frac{\beta(\frac{r}{5}+\frac{7}{10},\frac{r}{5}+\frac{7}{10})}{\beta(\frac{r}{5}+\frac{7}{10},-\frac{3r}{5}-\frac{1}{10})}
    },\quad r \ne 4\text{ mod } 5 \\
    &\quad \text{(the sign is $+$ for $r=0,3$ and $-$ for $r=1,2$ mod 5)}\\[2mm]
  A_1 \text{ series: }\qquad
  \frac{D_1}{D_0}&=16^{r+\half}\abs{\frac{\beta(\frac{r}{2}+\frac34,\frac{r}{2}+\frac34)}{\beta(\frac{r}{2}+\frac34,-\frac{3r}{2}-\frac14)}}\\[2mm]
  A_2 \text{ series: }\qquad
  \frac{D_1}{D_0}&=\frac{16^{\frac{2r+2}{3}}}{2}\abs{\frac{\beta(\frac{r}{3}+\frac56,\frac{r}{3}+\frac56)}{\beta(\frac{r}{3}+\frac56,1-r)}},\quad r \ne 2\text{ mod } 3 \\[2mm]
  D_4 \text{ series: }\qquad
  \frac{D_1}{D_0}&=\frac{16^{2r+1}}{3}\abs{\frac{\beta(r+1,r+1)}{\beta(r+1, -3r-1)}}
\end{split}
\ee

Thus we have determined the ratio of degeneracies for the quasi-characters (these were not explicitly determined in \cite{Chandra:2018pjq}). The first few nontrivial examples in each fusion class are presented in Appendix \ref{deg.2}.

\subsection{3 characters}

From \eref{cform}, the central charge and conformal dimensions of the candidate theory are given in terms of the parameters $a,\rho$ of the contour integral by:
\be
c=8(3a+\rho+1),\quad h_1=a+\half, \quad h_2=2a+\rho+1
\label{exponents.3}
\ee
and the normalisation constants are:
\begin{equation}
      \begin{split}
N_0 &= \frac{2!\, 16^{2a+\frac23 (\rho+1)}\,e^{-i\pi\rho}\,D_0}{S_2 (-1-3a-2\rho,a+1,\rho)} \frac{\mf s(\rho)}{\mf s(2\rho)} \\[2mm]
N_1 &=  \frac{\, 16^{\frac13 (2\rho-1)}D_1}{S_1(a+1,a+1,\rho)S_1 (-1-3a-2\rho,a+1,\rho)}  \\[2mm]
N_2 &= \frac{2!\, 16^{-2a-\frac43 (\rho+1)}\,e^{-i\pi\rho}\,D_2}{S_2 (a+1,a+1,\rho)}\frac{\mf s(\rho)}{\mf s(2\rho)} 
      \end{split}
      \label{norm.3}
    \end{equation}
where $S_2$ are the Selberg integrals defined and computed in \eref{Selberg}.

Next we compute the matrix $\hmS$ using our algorithm, verify that it squares to 1, and from it obtain the matrix $\mS$ via \eref{StoShat}. The result is:
\be
\mS=\begin{pmatrix}
\frac{\mf s(a)\mf s(a+\rho)}{\mf s(2a)\mf s(2a+\rho)} & 
-\frac{2e^{i\pi\rho}\mf s^2(a)\mf c(\rho)}{\mf s(2a)\mf s(2a+2\rho)}\frac{N_0}{N_1} & 
\frac{\mf s(a)\mf s(a+\rho)}{\mf s(2a+\rho)\mf s(2a+2\rho)}\frac{N_0}{N_2} \\[2mm]
-\frac{e^{-i\pi\rho}\mf s(a+\rho)\mf s(3a+\rho)}{\mf s(2a)\mf s(2a+\rho)} \frac{N_1}{N_0}& 
~~\frac{\mf s(a)\mf s(3a+\rho)}{\mf s(2a)\mf s(2a+\rho)} 
-\frac{\mf s(a)\mf s(a+\rho)}{\mf s(2a+\rho)\mf s(2a+2\rho)}~~ 
& 
\frac{e^{-i\pi\rho}\mf s^2(a+\rho)}{\mf s(2a+\rho)\mf s(2a+2\rho)}\frac{N_1}{N_2} \\[2mm]
\frac{\mf s(3a+\rho)\mf s(3a+2\rho)}{\mf s(2a)\mf s(2a+\rho)} \frac{N_2}{N_0}& 
\frac{2e^{i\pi\rho}\mf s(a)\mf s(3a+2\rho)\mf c(\rho)}{\mf s(2a)\mf s(2a+2\rho)}\frac{N_2}{N_1} & 
\frac{\mf s(a)\mf s(a+\rho)}{\mf s(2a+\rho)\mf s(2a+2\rho)}
\end{pmatrix}
\label{Smat.3}
\ee
Though it may not be immediately apparent, this matrix is in fact real. The reason is that the phases coming from $\hmS$ are exactly cancelled by corresponding phases in the normalisation factors in \eref{norm.3}\footnote{In subsequent sections we will see that this property holds in all our examples, though we did not yet find a general proof of it.}. 

The general question now is to find the solutions of:
\begin{equation}
  \mathcal{S}^\dagger \begin{pmatrix} 1 & 0 &0\\ 0 & M_1 & 0\\
0 & 0 & M_2  
  \end{pmatrix} \mathcal{S} = \begin{pmatrix} 1 & 0 &0\\ 0 & M_1 & 0\\
0 & 0 & M_2  
\end{pmatrix}
\label{SMatProp.3}
\end{equation}
for various possible positive integers $M_i$.  Imposing this requirement leads to:
 
\be
\begin{split}
\frac{N_1}{N_0} &=\pm \frac{e^{i \pi \rho}}{\sqrt{M_1}}\sqrt\frac{\mfs(2\rho)\,\mfs^2(a)\,\mfs(2a+\rho)}{\mfs(\rho)\,\mfs(a+\rho)\,\mfs(2a+2\rho)\,\mfs(3a+\rho)}\\[2mm]
\frac{N_2}{N_0} &=\pm \frac{1}{\sqrt{M_2}}\sqrt\frac{\mfs(a)\,\mfs(2a)\,\mfs(a+\rho)}{\mfs(2a+2\rho)\,\mfs(3a+\rho)\,\mfs(3a+2\rho)}
\end{split}
\label{nratios.3}
\ee

There is a nice simplification if we consider the case where all multiplicities are equal to 1. Then \eref{SMatProp.3} says that $\mS$ is unitary: $\mS^\dagger \mS=1$. But 
we also know that $\mS^2=1$. It follows that $\mS$ must be Hermitian. But since the matrix in \eref{Smat.3} is real, it must also be symmetric. Hence to ensure unitarity of $\mS$ we only need to impose the requirement of symmetry. This easily gives us \eref{nratios.3} for the special case $M_1=M_2=1$.

The next step is to set $M_1=M_2=1$ and use the Verlinde formula and integrality of the fusion rule coefficients to constrain possible theories. This time we will use the result of \cite{Christe:1988xy} where possible fusion classes for theories with three primaries (including the identity) were classified. There are altogether four such classes, of which three have the fusion rules:
\be
\begin{split}
& {\cal A}_2^{(1)}:\quad \cN_{011}=\cN_{022}=\cN_{122}=1\\
& {\cal A}_2^{(2)}:\quad \cN_{011}=\cN_{022}=\cN_{122}=\cN_{222}=1\\
& {\cal A}_2^{(3)}:\quad \cN_{011}=\cN_{022}=\cN_{122}=\cN_{112}=\cN_{222}=1
\end{split}
\ee
where the remaining fusion rule coefficients are zero. We have not written the fourth class ${\cal B}_2^{(1)}$ because although it describes three primaries, the theory has only two characters (the two non-trivial primaries are complex conjugates of each other). Thus this class is not relevant for us here.

Now from the Verlinde formula we have:
\be
\cN_{111}=\frac{\mS_{10}^3}{\mS_{00}}+\frac{\mS_{11}^3}{\mS_{01}}+\frac{\mS_{12}^3}{\mS_{02}}
\ee
Plugging in the above $\mS$-matrix, we get:
\be
\begin{split}
\cN_{111}&=\half\frac{N_1}{N_0}\frac{e^{-3\pi i \rho}\,\mfs(a)\,\mfs(2a+2\rho)}{\mfs^2(2a)\,\mfs^2(2a+\rho)}\Bigg[ \left(\frac{\mfs(a+\rho)}{\mfs(2a+2\rho)}-\frac{\mfs(3a+\rho)}{\mfs(2a)}\right)^3\frac{e^{2\pi i \rho}\mfs^3(2a)}{\mfc(\rho)\mfs(2a+\rho)}\\[1mm]
&\qquad -2\, \frac{N_1^2}{N_0^2}\,\frac{\mfs^2(a+\rho)\,\mfs^3(3a+\rho)}{\mfs^2(a)\,\mfs(2a+2\rho)}
+2\, \frac{N_1^2}{N_2^2}\,\frac{\mfs^5(a+\rho)\,\mfs^2(2a)}{\mfs^2(a)\,\mfs^3(2a+2\rho)}\Bigg]
\end{split}
\label{N111constraint}
\ee
As a check of this formula, we input the values $a=-\frac{11}{14}$ and $\rho=\frac17$ corresponding to the ${\cal M}_{2,7}$ non-unitary minimal model and find $\cN_{111}=0$. Similarly the remaining fusion rules are found to be $\cN_{222}=0, \cN_{112}=\cN_{122}=1$. This corresponds to 
the class ${\cal A}_2^{(3)}$. Similarly we can verify that SU(2) at level 2 as well as ${\cal M}_{3,4}$ (the Ising CFT) are both in the class ${\cal A}_2^{(1)}$.

The remaining $\cN_{ijk}$ are likewise rather complicated functions of $a,
\rho$. The question then arises whether we can use \eref{N111constraint} to classify possible three-character theories rather than starting with the $a,\rho$ values of a known theory. Unfortunately this seems a rather daunting task. So we will turn to other applications.

First of all we obtain explicit formulae for the degeneracies of primary fields in these theories. Combining \eref{norm.3} and \eref{nratios.3} we find:
\be
\begin{split}
\frac{D_1}{D_0} &= \frac{16^{2a+\frac54}}{\sqrt{M_1}}
\abs{
\frac{S_1(-1-3a-2\rho,a+1,\rho)S_1(a+1,a+1,\rho)}{S_2(-1-3a-2\rho,a+1,\rho)}}\times\\
&\qquad \qquad \qquad 
\abs{\frac{\mfs(\rho)\,\mfs^2(a)\,\mfs(2a+\rho)}{\mfs(2\rho)\,\mfs(a+\rho)\,\mfs(2a+2\rho)\,\mfs(3a+\rho)}}^\half\\[2mm]
\frac{D_2}{D_0} &= \frac{16^{4a+2\rho+2}}{\sqrt{M_2}}
\abs{\frac{S_2(a+1,a+1,\rho)}{S_2(-1-3a-2\rho,a+1,\rho)}
}\times \\[2mm]
&\qquad\qquad\qquad
\abs{\frac{\mfs(a)\,\mfs(2a)\,\mfs(a+\rho)}{\mfs(2a+2\rho)\,\mfs(3a+\rho)\,\mfs(3a+2\rho)}}^\half
\end{split}
\label{Dratios.3}
\ee

We may now apply this to various known or predicted theories. We will skip the step of applying them to $c<1$ minimal models, because all degeneracies there are equal to 1, and WZW models because all degeneracies are given by the dimensions of Lie algebra representations (one can of course re-derive these facts using the above formulae). So we will instead consider a more exotic family of three-character theories, namely the novel cosets predicted in \cite{Gaberdiel:2016zke}. In Table \ref{Degen.3} we list the degeneracies $D_1,D_2$ and multiplicities $M_1,M_2$ for each of the three-character cosets in that  paper. Wherever $M_1=M_2=1$, we also checked the fusion class and found that for all entries in the table, they correspond to ${\cal A}_2^{(1)}$.

\begin{table}
  \centering
  \begin{tabular}{ccccccc}
    \toprule
 $c$ & $h_1$ & $h_2$ & $D_1$ & $D_2$ & $M_1$ & $M_2$ \\
    \midrule
      $\frac{45}{2}$ & $\frac{29}{16}$ & $\frac32$ & 46080 & 4785 & 1 & 1  \\[2mm]
            $\frac{43}{2}$ & $\frac{27}{16}$ & $\frac32$ & 22016 & 5031 & 1 & 1 \\[2mm]
  $21$ & $\frac{13}{8}$ & $\frac32$ & 10752 & 5096 & 2 & 1 \\[2mm]
    $\frac{41}{2}$ & $\frac{25}{16}$ & $\frac32$ & 10496 & 5125 & 1 & 1 \\[2mm]
      $20$ & $\frac{8}{5}$ & $\frac75$ & 8125 & 2500 & 2 & 2 \\[2mm]
        $\frac{39}{2}$ & $\frac{23}{16}$ & $\frac32$ & 4992 & 5083 & 1 & 1 \\[2mm]
          $19$ & $\frac{11}{8}$ & $\frac32$ & 2432 & 5016 & 2 & 1 \\[2mm]
      $\frac{37}{2}$ & $\frac{21}{16}$ & $\frac32$ & 2368 & 4921 & 1 & 1  \\[2mm]
      $18$ & $\frac{5}{4}$ & $\frac32$ & 1152 & 4800 & 2 & 1 \\[2mm]
  $\frac{35}{2}$ & $\frac{19}{16}$ & $\frac32$ & 1120 & 4655 & 1 & 1 \\[2mm]
    $17$ & $\frac{9}{8}$ & $\frac32$ & 544 & 4488 & 2 & 1 \\[2mm]
      $\frac{31}{2}$ & $\frac{15}{16}$ & $\frac32$ & 248 & 3875 & 1 & 1 \\[2mm]
        $\frac{17}{2}$ & $\frac{17}{16}$ & $\frac12$ & 256 & 17 & 1 & 1 \\[2mm]
          $15$ & $\frac{7}{8}$ & $\frac32$ & 120 & 3640 & 2 & 1\\[2mm]
   $14$ & $\frac{3}{4}$ & $\frac32$ & 56 & 3136 & 2 & 1\\                
    \bottomrule
  \end{tabular}
  \caption{Degeneracies for 3-character coset theories. Wherever $M_1=M_2=1$, the fusion class is also provided.}
  \label{Degen.3}
\end{table}

\newpage

\subsection{4 characters}

%There is an interesting subtlety that arises in the context of contour integral representations. The fusion rules as written above are not symmetric under the interchange $1\leftrightarrow 2$. In \cite{Christe:1988xy} it was found that $N_{\phi\phi\phi}=0$ for at least one field, which was then labelled 1. However in the contour representation there is already a definite choice of the field labelled 1 (the two contours run over different regions $1\to\infty$ and $0\to\lambda$), and the field labelled 2 (the two contours run over the same region, $0 \to \lambda$). Thus given a theory with a contour-integral representation, it could either match the above fusion rules as written, or an identical set with the role of 1 and 2 interchanged. We will refer to the latter as ${\cal A'}_2^{(1)}, {\cal A'}_2^{(2)}, {\cal A'}_2^{(3)}$.

We now move on to the case of 4 characters. The $\mS$-matrix for this case has not previously appeared in the literature. A new feature with 4 characters is that one can obtain such a set in two distinct ways: $(n_1,n_2)=(3,0)$ and $(n_1,n_2)=(1,1)$ where we recall that $n_1,n_2$ is the number of integration variables of type $t_i$ and $\tau_j$ respectively in \eref{calJdef}. However, integrals with $n_2\ne 0$ have not been addressed in this paper for simplicity. For those with $(n_1,n_2)=(3,0)$, the $\mS$-matrix and normalisations are readily computable using the algorithm developed here. We find the normalisations to be:
\be
\begin{split}
  N_0 &= \frac{3!\, 16^{1+3a+2\rho}\,e^{-i\pi3\rho}\,D_0}{S_3 (-1-3a-4\rho,a+1,\rho)} \frac{\mf s^2(\rho)}{\mf s(2\rho)\mf s(3\rho)} \\[2mm]
  N_1 &= \frac{2!\, 16^{a+2\rho}\,e^{-i\pi\rho} D_1}{S_1(a+1,a+1,\rho)S_2 (-1-3a-4\rho,a+1,\rho)}\frac{\mf s(\rho)}{\mf s(2\rho)}  \\[2mm]
  N_2 &= \frac{2!\, 16^{1+a}\,e^{-i\pi\rho} D_2}{S_2(a+1,a+1,\rho)S_1 (-1-3a-4\rho,a+1,\rho)} \frac{\mf s(\rho)}{\mf s(2\rho)}   \\[2mm]
  N_3 &= \frac{3!\, 16^{2+3a+4\rho}\,e^{-i\pi3\rho}\,D_3}{S_3 (a+1,a+1,\rho)}\frac{\mf s^2(\rho)}{\mf s(2\rho)\mf s(3\rho)}
\end{split}
\label{norm.4}
\ee
Next we list the $\mS$ matrix. Because of the width of the columns, we have separately displayed each column.
\be
\begin{gathered}
\mc S_{A0} = 
\begin{pmatrix}
  \frac{\mf s(a) \mf s(a+\rho) \mf s(a+2\rho)}{\mf s(2a) \mf s(2 a+\rho) \mf s(2 a+2 \rho) }  \\[2mm]
  -e^{-2i\pi\rho }\frac{\mf s(a+\rho)\mf s(a+2\rho)\mf s(3a+2\rho)}{\mf s(a)\mf s(2a+\rho) \mf s(2a+2\rho)}\frac{N_1}{N_0}\\[2mm]
  e^{-2i \pi \rho }\frac{\mf s(a+2 \rho) \mf s(3 a+2 \rho) \mf s(3 a+3 \rho)}{ \mf s(2 a ) \mf s(2 a+\rho) \mf s(2 a+2 \rho)}\frac{N_2}{N_0}\\[2mm]
  -\frac{\mf s(3 a+2 \rho) \mf s(3 a+3 \rho) \mf s(3 a+4 \rho)}{\mf s(2 a) \mf s(2 a+\rho) \mf s(2 a+2 \rho) }\frac{N_3}{N_0}
\end{pmatrix} \\[2mm]
\mc S_{A1} = 
\begin{pmatrix}
 - e^{2i\pi\rho}\left(
  \frac{\mf s(a+\rho) \mf s^2(a)}{\mf s(2 a+\rho) \mf s^2(2 a+2 \rho ) }
  +\frac{\mf s(a+\rho) \mf s^2(a)}{\mf s(2 a) \mf s(2 a+\rho) \mf s(2 a+2 \rho)} 
  +\frac{ \mf s(a+\rho ) \mf s^2(a) }{\mf s^2(2 a+2 \rho) \mf s(2 a+3 \rho)}\right)\frac{N_0}{N_1}\\[2mm]
  -\frac{\mf s(a) \mf s(a+\rho) \mf s(a+2\rho)}{\mf s^2(2 a+2 \rho)\mf s(2 a+3 \rho)}
  +\frac{\mf s(a) \mf s(a+\rho) \mf s(3 a+2 \rho)}{\mf s(2 a+\rho)\mf s^2(2 a+2 \rho)}
  +\frac{\mf s(a) \mf s(a+\rho) \mf s(3 a+2 \rho)}{\mf s(2 a) \mf s(2 a+\rho)\mf s(2 a+2 \rho)}\\[2mm]
  \left(\frac{ \mf s(3 a+3 \rho) \mf s^2(a+\rho)}{\mf s(2 a+\rho) \mf s^2(2 a+2 \rho) }
  +\frac{\mf s(3 a+3 \rho) \mf s^2(a+\rho)}{\mf s^2(2 a+2 \rho) \mf s(2 a+3 \rho) }
  -\frac{ \mf s(a ) \mf s(3 a+2 \rho) \mf s(3 a+3 \rho)}{\mf s(2 a ) \mf s(2 a+\rho) \mf s(2 a+2 \rho)}\right)\frac{N_2}{N_1}\\[2mm]
   -e^{2i \pi \rho }\left(
   \frac{\mf s(a) \mf s(3 a+3 \rho) \mf s(3 a+4 \rho)}{\mf s(2 a+\rho) \mf s^2(2 a+2 \rho)} 
  +\frac{\mf s(a) \mf s(3 a+3 \rho) \mf s(3 a+4 \rho)}{\mf s(2 a+3 \rho)\mf s^2(2 a+2 \rho)}
  +\frac{\mf s(a) \mf s(3 a+3 \rho) \mf s(3 a+4 \rho)}{\mf s(2 a) \mf s(2 a+\rho) \mf s(2 a+2 \rho)}
  \right)\frac{N_3}{N_1}
\end{pmatrix} \\[2mm]
\mc S_{A2} = 
\begin{pmatrix}
  e^{2i\pi\rho} \left(
  \frac{\mf s(a+\rho ) \mf s^2(a)}{\mf s(2 a+\rho) \mfs^2(2a+2\rho)} 
  +\frac{\mf s(a+\rho) \mf s^2(a)}{ \mfs^2(2 a+2 \rho) \mf s(2 a+3 \rho)} 
  +\frac{ \mf s(a+\rho) \mf s^2(a)}{\mfs(2a+2\rho)\mf s(2 a+3 \rho) \mf s(2 a+4 \rho)}\right)\frac{N_0}{N_2}\\[2mm]
  \left(-\frac{\mf s(a) \mf s(a+\rho) \mf s(3 a+2 \rho) }{\mf s(2 a+\rho)\mf s^2(2 a+2 \rho)} 
  +\frac{ \mf s(a) \mf s(a+\rho) \mf s(a+2 \rho) }{\mf s(2 a+3 \rho)\mf s(2 a+2 \rho) \mf s(2 a+4 \rho)}
  +\frac{ \mf s(a) \mf s(a+\rho) \mf s(a+2 \rho)}{\mf s^2(2 a+2 \rho) \mf s(2 a+3 \rho)}\right)\frac{N_1}{N_2} \\[2mm]
 - \frac{ \mf s(3 a+3 \rho) \mf s^2(a+\rho)}{\mf s(2 a+\rho) \mf s^2(2 a+2 \rho)}
  -\frac{\mf s(3 a+3 \rho) \mf s^2(a+\rho)}{\mf s^2(2 a+2 \rho) \mf s(2 a+3 \rho)} 
  +\frac{\mf s(a ) \mf s(a+2 \rho) \mf s(a+\rho)}{\mf s(2 a+2 \rho) \mf s(2 a+3 \rho) \mf s(2 a+4 \rho)}\\[2mm]
  - e^{2 i \pi  \rho } \left(
  \frac{ \mf s(a) \mf s(a+\rho) \mf s(3 a+4 \rho)}{\mf s(2 a+\rho) \mf s^2(2 a+2 \rho)}
  +\frac{ \mf s(a) \mf s(a+\rho) \mf s(3 a+4 \rho) }{\mf s(2 a+3 \rho)\mf s^2(2 a+2 \rho)} 
  +\frac{\mf s(a) \mf s(a+\rho) \mf s(3 a+4 \rho) }{\mf s(2 a+3 \rho) \mf s(2 a+4 \rho)\mf s(2 a+2 \rho) }
  \right)\frac{N_3}{N_2}
\end{pmatrix}\\[2mm]
\mc S_{A3} = 
\begin{pmatrix}
 - \frac{\mf s(a) \mf s(a+\rho) \mf s(a+2 \rho)}{\mf s(2 a+2 \rho) \mf s(2 a+3 \rho) \mf s(2 a+4 \rho) } \frac{N_0}{N_3}\\[2mm]
  -e^{-2i \pi\rho }\frac{\mf s(a+\rho) \mf s^2(a+2 \rho) }{ \mf s(2 a+2 \rho) \mf s(2 a+3 \rho) \mf s(2 a+4 \rho)}\frac{N_1}{N_3}\\[2mm]
 - e^{-2i \pi\rho }\frac{\mf s(a+\rho) \mf s^2(a+2 \rho)}{ \mf s(2 a+2 \rho) \mf s(2 a+3 \rho) \mf s(2 a+4 \rho)}\frac{N_2}{N_3}\\[2mm]
 - \frac{\mf s(a) \mf s(a+\rho) \mf s(a+2 \rho)}{\mf s(2 a+2 \rho) \mf s(2 a+3 \rho) \mf s(2 a+4 \rho)}\\[2mm]
\end{pmatrix}
\end{gathered}
\ee
Notice that each entry in the second and third columns is made up of a sum of three terms. This reflects the fact that there are three ``paths'' in \eref{coolgraph} to reach the desired point. On the other hand, the entries in the first and fourth columns have only one term each, because for those cases there is a unique path.

As explained at the outset, with four characters there is no general guarantee that a contour integral representation exists for some given characters (with $\ell=0)$. However if the critical exponents of a given set of characters can be expressed in terms of some parameters $a,\rho$ via \eref{exponents.3} then indeed the contour integrals with those values of $a,\rho$ must uniquely give those characters. We will apply the above formulae to five such theories with central charges $\frac95, 5, \frac{21}{5},\frac{14}{3},\frac{58}{3}$ corresponding respectively to the WZW models $A_{1,3}$, $A_{5,1}$, $C_{3,1}$, $G_{2,2}$ and the coset dual of $G_{2,2}$ found in Table 3 of \cite{Gaberdiel:2016zke}. The results are given in Table \ref{Degen.4} where we label these theories by their central charges. We see, as always, the curious fact that coset theories have large degeneracies.

\begin{table}
  \centering
  \begin{tabular}{cccccccccc}
    \toprule
 $c$ & $h_1$ & $h_2$ & $h_3$ & $D_1$ & $D_2$ & $D_3$ & $M_1$ & $M_2$ & $M_3$\\
    \midrule
      $\frac{9}{5}$ & $\frac{3}{20}$ & $\frac25$ & $\frac34$ & 2 & 3 & 4 & 1 & 1 & 1\\[2mm]
            5 & $\frac{5}{12}$ & $\frac23$ & $\frac34$ & 6 & 15 & 20 & 2 & 2 & 1\\[2mm]
  $\frac{21}{5}$ & $\frac{7}{20}$ & $\frac35$ & $\frac34$ & 6 & 14 & 14 & 1 & 1 & 1\\[2mm]
    $\frac{14}{3}$ & $\frac{7}{9}$ & $\frac13$ & $\frac23$ & 7 & 14 & 27 & 1 & 1 & 1\\[2mm]
        $\frac{58}{3}$ & $\frac{11}{9}$ & $\frac53$ & $\frac43$ & 1044 & 16588 & 1595 & 1 & 1 & 1\\                 
    \bottomrule
  \end{tabular}
  \caption{Degeneracies for 4-character coset theories}
  \label{Degen.4}
\end{table}

\section{Greater than 5 characters}

As an example of the power of our algorithm, we can speedily compute the matrix $\hmS$ for the WZW model of SU(2) at  any level $k$, a theory with $k+1$ characters. For these theories the parameters $a,\rho$ are:
\be
a=-\frac{2k+1}{4(k+2)},\quad \rho=\frac{1}{2(k+2)}
\ee
It was conjectured in \cite{Mukhi:1989qk} that the contour integral with $n=k$  integration variables and these values for the parameters correctly describes the $k+1$ characters of SU(2)$_k$. For $n\le 4$ this is guaranteed by the fact that the exponents of the theory can be parametrised in terms of $a,\rho$, because with this data one can write an MLDE that is uniquely determined. However for $n>4$ the MLDE is no longer unique, thus this is still  a conjecture -- for which some evidence exists. If we could calculate the entire $\mS$-matrix in closed form for all $k$ using our method we would be able to compare it with the known $\mS$-matrix for the WZW model:
\be
\mS_{jj'}=\sqrt\frac{2}{k+2}~\sin\frac{(2j+1)(2j'+1)\pi}{k+2}
\label{kacweyl}
\ee
While this seems difficult to do in general, it is very easy in any specific case. For example at $k=7$ our method gave rise to an $\mS$-matrix whose numerical values are:
\be
\begin{pmatrix}
  0.1612 & 0.3030 & 0.4082 & 0.4642 & 0.4642 & 0.4082 & 0.3030 & 0.1612 \\
  0.3030 & 0.4642 & 0.4082 & 0.1612 & -0.1612 & -0.4082 & -0.4642 & -0.3030 \\
  0.4082 & 0.4082 & 0 & -0.4082 & -0.4082 & 0 & 0.4082 & 0.4082 \\
  0.4642 & 0.1612 & -0.4082 & -0.3030 & 0.3030 & 0.4082 & -0.1612 & -0.4642 \\
  0.4642 & -0.1612 & -0.4082 & 0.3030 & 0.3030 & -0.4082 & -0.1612 & 0.4642 \\
  0.4082 & -0.4082 & 0 & 0.4082 & -0.4082 & 0 & 0.4082 & -0.4082 \\
  0.3030 & -0.4642 & 0.4082 & -0.1612 & -0.1612 & 0.4082 & -0.4642 & 0.3030 \\
  0.1612 & -0.3030 & 0.4082 & -0.4642 & 0.4642 & -0.4082 & 0.3030 & -0.1612 \\
\end{pmatrix}
\ee
It is easily verified that term by term, this matches the formula \eref{kacweyl}. This goes beyond the checks in \cite{Mukhi:1989qk} where only the last row of the matrix was computed and confirmed.

We may mention here that the first and last rows of $\mS$ are easy to compute for a reason: in our algorithm \eref{coolgraph} there is a unique path that leads from the vertex at the top to either the left-most or right-most vertex at the bottom. Thus the answer is a straightforward product of simple factors. We find:
\be
\begin{split}
S_{A0}&=(-1)^n \frac{N_A}{N_0} \prod_{s=0}^{A-1}\frac{e^{i\pi(A-2s-1)\rho}\,\mfs(3a+(n+A-s-2)\rho)}{\mfs(2a+(n-s-1)\rho)} \prod_{s=A}^{n}\frac{-e^{i\pi(n-2s-1)\rho}\,\mfs(a+s\rho)}{\mfs(2a+(n-s-1)\rho)}\\
S_{An}&=(-1)^n \frac{N_A}{N_n} \prod_{s=0}^{A-1}\frac{e^{i\pi(A-2s-1)\rho}\,\mfs(a+(n-A+s)\rho)}{\mfs(2a+(n+s-1)\rho)} \prod_{s=A}^{n}\frac{e^{i\pi(n-2s-1)\rho}\,\mfs(a+s\rho)}{\mfs(2a+(n+s-1)\rho)}
\end{split}
\ee

\section{Conclusions and open problems}

We have systematised and provided additional evidence for the proposal of \cite{Mukhi:1989qk} to describe the characters of RCFT via contour integrals. Our algorithm to compute the auxiliary matrix $\hmS$ has an elegant representation as a sum over paths, while the normalisations are computed using the Selberg integral formula. A technical limitation of the present paper is that we only considered integrals with one set of variables $t_i$ rather than the most general case which has two sets $t_i,\tau_j$. We hope to address the latter case in the future. We now describe some interesting directions provided by this work. 

For cases with unit multiplicity, the Verlinde formula applied to $\mS$ provides constraints on the parameters of the contour integral, which in principle could be used to classify RCFT with arbitrary numbers of characters (and vanishing Wronskian index). Unfortunately this already proved somewhat intractable from three characters onwards, but this technical problem might be possible to overcome. 

The conjecture of \cite{Mukhi:1989qk} is still unproven in general. However we were able to extend it to coset theories with 3,4 characters which had not been discovered at the time.
We also added some evidence to the conjecture here, but much more can be done. One of the nice goals would be to reproduce the simple form of $\mS$ for SU(2)$_k$ WZW theories for all $k$. In our approach it should arise by summing over paths.

In addition to describing RCFT, we saw that pairs of contour integrals can describe quasi-characters \cite{Chandra:2018pjq}. These are powerful building blocks for generic two-character RCFT. The analogous construction has not yet been carried out for three or more characters. The contour integral representation provides a useful way to approach the problem, when coupled with a study of the $q$-series. We hope to report on this in the future.

\section*{Acknowledgements}

RP and PS would like to thank Prasanna Joshi and Raj Patil for helpful discussions. They also acknowledge the INSPIRE Scholarship for Higher Education, Government of India. All of us are grateful for support from a grant by Precision Wires India Ltd.\ for String Theory and Quantum Gravity research at IISER Pune.

\section*{Appendices}

\appendix

\section{Review of the ``unfolding'' procedure}
\label{unfolding}
 
Here we illustrate the ``unfolding'' procedure that allows one to compute the modular $S$-transformation. This uses contour deformation methods and we will work it out in the simplest case $(n_1,n_2)=(1,0)$. 
In this case the characters are:
 \be
J_A(\lambda)=N_A \left[ \lambda (1-\lambda) \right]^\alpha \hJ_A(\lambda), ~A=0,1
\label{IIhat.2}
  \ee
where
\be
\begin{split}
\hJ_0(\lambda)&=\int_1^\infty dt
\Big[t(t-1)(t-\lambda)\Big]^a\\
\hJ_1(\lambda)&= \int_0^\lambda dt
\Big[t(1-t)(\lambda-t)\Big]^a
\end{split}
\ee
Now we compute the effect of the modular $\mS$-transformation $\lambda\to 1-\lambda$ on these integrals. We start with $\hJ_0$ and deform the contour in a counter-clockwise manner to go from $1$ to $-\infty$. This is justified because the integral over a semi-circular contour in the upper half-plane is zero (it encloses no singular points) and the contribution from the arc at infinity also contributes zero (for values of the parameters for which the integrand falls off at infinity, which is the case when the calculations are done -- later we extend by analytic continuation).
Thus we have:
\be
\begin{split}
\hJ_0 &= e^{\pi i a}\int_{1^+}^{\lambda^+} dt \Big[t(1-t)(t-\lambda)\Big]^a +e^{2\pi ia}\int_{\lambda^+}^{0^+} dt\Big[t(1-t)(\lambda-t)\Big]^a\\
&\quad + e^{3\pi ia}\int_{0^+}^{-\infty} dt \Big[(-t)(1-t)(\lambda-t)\Big]^a
\end{split}
\label{pluscon}
\ee
Here by $1^+,\lambda^+,0^+$ we mean that the contours go around these singular points with a small positive imaginary part. Notice that in each segment, the integrand has been re-ordered so that all brackets are positive, and their the fractional powers are defined to have no phase. The explicit phase that arises is written outside the integral.  

Next we perform a similar deformation of the original contour clockwise (because a semi-circular contour in the lower half-plane also gives a vanishing contribution, by the same arguments), to get:
\be
\begin{split}
\hJ_0 &= e^{-\pi i a}\int_{1^-}^{\lambda^-} dt \Big[t(1-t)(t-\lambda)\Big]^a +e^{-2\pi ia}\int_{\lambda^-}^{0^-} dt\Big[t(1-t)(\lambda-t)\Big]^a\\
&\quad + e^{-3\pi ia}\int_{0^-}^{-\infty} dt \Big[(-t)(1-t)(\lambda-t)\Big]^a
\end{split}
\label{minuscon}
\ee
where this time the contours go around the singular points with a negative imaginary part. 

Multiplying \eref{pluscon} by $e^{-2\pi ia}$ and \eref{minuscon} by $e^{2\pi i a}$ and subtracting the second from the first, we have:
\be
\begin{split}
-2i\sin 2\pi a~ \hJ_0 &=-2i\sin\pi a \int_{1}^{\lambda} dt \Big[t(1-t)(t-\lambda)\Big]^a\\
&\qquad + 2i\sin\pi a \int_{0}^{-\infty} dt \Big[(-t)(1-t)(t-\lambda)\Big]^a\\
&=2i\sin\pi a \int_{0}^{1-\lambda} dt \Big[t(1-t)(1-\lambda-t)\Big]^a\\
&\qquad -2i\sin\pi a \int_{1}^{\infty} dt \Big[t(t-1)(t-1+\lambda)\Big]^a
\end{split}
\label{elimseg}
\ee
In the first line above, notice that the segment between 0 and $\lambda$ has dropped out.

Recognising that the integrals on the RHS are the same $\hJ_A$ but with argument $1-\lambda$, we find:
\be
\hJ_0(\lambda)=\frac{\sin\pi a}{\sin 2\pi a}\big(J_0(1-\lambda)-J_1(1-\lambda)\big)
\ee
In a similar way we have:
\be
\begin{split}
\hJ_1(\lambda)&=e^{\pi i a}\int_{0^+}^{-\infty}\Big[(-t)(1-t)(\lambda-t)\Big]^a + e^{-2\pi ia}\int_\infty^{1^+}\Big[t(t-1)(t-\lambda)\Big]^a\\
&\quad +e^{-\pi ia}\int_{1^+}^{\lambda^+}\Big[t(1-t)(t-\lambda)\Big]^a\\
&=e^{-\pi i a}\int_{0^-}^{-\infty}\Big[(-t)(1-t)(\lambda-t)\Big]^a+ e^{2\pi ia}\int_\infty^{1^-}\Big[t(t-1)(t-\lambda)\Big]^a\\
&\quad +e^{\pi ia}\int_{1^-}^{\lambda^-}\Big[t(1-t)(t-\lambda)\Big]^a
\end{split}
\ee
From this we read off that:
\be
\hJ_1(\lambda) = -\frac{\sin 3\pi a}{\sin 2\pi a}\hJ_0(1-\lambda)
 -\frac{\sin \pi a}{\sin 2\pi a}\hJ_1(1-\lambda)
\ee
Thus, we have shown that:
\be
\hJ_A(\lambda)=\sum_B \hS_{AB}\hJ_B(1-\lambda)
\label{Ihattrans}
\ee
where:
\be
\hS_{AB}=\begin{pmatrix}
\frac{\sin\pi a}{\sin 2\pi a} & ~~-\frac{\sin\pi a}{\sin 2\pi a}\\[3mm]
-\frac{\sin 3\pi a}{\sin 2\pi a} & ~~-\frac{\sin\pi a}{\sin 2\pi a}
\end{pmatrix}
\label{hatS}
\ee

This example has been deceptively simple and is presented only to illustrate the basic idea. The key points illustrated here is that after deforming contours, the resulting contributions from contours in the segments $(0,-\infty)$ and $(\lambda, 1)$ are retained and after sending $t\to 1-t$ they correspond to the original characters as functions of modular transformed variable $1-\lambda$. On the other hand, any contributions from the regions $(\lambda,0)$ and $(1,\infty)$ obtained after contour deformation need to be eliminated by taking a suitable linear combination, as in \eref{elimseg}, because the integral in these regions does not map to any of the original characters after $t\to 1-t$.

Things become much more complicated when we have two or more contours. The factors $(t_i-t_k)^{2\rho}$, absent in the one-variable case, are responsible for the complications. Whenever the $t_i$ and $t_k$ contours run over the same region, the bracketed factors change sign even within the same region, according to whether $t_i>t_k$ or the other way around. Following the original works, we choose the convention that the bracketed factors have positive imaginary parts, i.e. when $\Im(t_i)>\Im(t_k)$ then we write $(t_i-t_k)$.  This introduces an asymmetry between the different $t_i$ and consequently an asymmetry between the operations of deforming a contour in the upper and lower half-planes. Thus the $\rho$-dependent phases will need more care to compute than the $a$-dependent phases. This is done in the main body of the paper.

\section{Relation between ordered and unordered integrals}
\label{JtoI_App}

In this appendix we prove the relation Eq.\ \eqref{oneparamIdef}, the relation between unordered integrals $J_A$ and ordered integrals $I_A$. First we will prove it for the case of two integration variables and then generalise it to give a proof by induction for an arbitrary number of variables.

We start by noting that $J_0$ is of the general form:
\be
J_0= \int_a^b dt_2 \int_a^b dt_1\, (t_2 - t_1)^{2\rho} f(t_1,t_2)
\ee 
where $f(t_1,t_2)$ is a symmetric function of its arguments. Now we can break up the integral of $t_1$ into two regions, corresponding to $t_1$ greater and less than $t_2$, to get:
\be
\begin{split}
 J_0 &=\int_a^b dt_2 \int_a^{t_2} dt_1\, (t_2 - t_1)^{2\rho}f(t_1,t_2) + \int_a^b dt_2 \int_{t_2}^b dt_1\, (t_2 - t_1)^{2\rho} f(t_1,t_2) \\[2mm]
  &= \int_a^b dt_2 \int_a^{t_2} dt_1 \,(t_2 - t_1)^{2\rho} f(t_1,t_2)+ e^{2i\pi\rho} \int_a^b dt_2 \int_{t_2}^b dt_1\, (t_1 - t_2)^{2\rho} f(t_1,t_2)\\[2mm]
\end{split}
\ee 
In the second term, to keep the integrand real, we have pulled out the phase $e^{2i\pi\rho}$. The sign in the exponent is determined by the fact that the imaginary parts of the contours were ordered to start with. By a re-labelling of the second integral we get:
\be
\begin{split}
J_0 &= \big(1+e^{i\pi2\rho}\big)  \int_a^b dt_2 \int_a^{t_2} dt_1 (t_2 - t_1)^{2\rho} f(t_1,t_2)\\
&= e^{i\pi\rho}\,\frac{\mfs(2\rho)}{\mfs(\rho)}\,I_0
\end{split}
\ee
This proves the relation for the case of $A=0$ and two integration variables (it trivially extends to $A=n$, and later we will extend it to the case where $A$ lies between 0 and $n$ which means the integration regions are partitioned between two different ranges). 

By considering more integration variables successively one easily guesses that the proportionality factor between the ordered integrals $I_0$ and unordered integrals $J_0$ is:
\be
\prod_{k=1}^{n-1} \sum_{m=0}^k e^{2\pi i\,m\rho}
\ee
To prove the general result for $n$ integration variables (but still $A= 0$ or $n$), we start with the following integral, and assume the above result for $n-1$ integrals, leading to a proof by induction: 
\be
\begin{split}
  &\quad \int_a^b dt_n \cdots  \int_a^b dt_1 f(t_i) \prod_{n\geq j > i \geq 1} (t_j - t_i)^{2\rho} \\
  &= \left(\prod_{k=1}^{n-2} \sum_{m=0}^k e^{2\pi i\,m\rho}\right) \int_a^b dt_n \int_a^b dt_{n-1}\int_a^{t_{n-1}} dt_{n-2} \cdots \int_a^{t_2} dt_1 \prod_{n\geq j > i \geq 1} f(t_i) (t_j - t_i)^{2\rho}
\end{split}
\ee
where $f(t_i)$ is totally symmetric in its arguments. Now we observe that splitting the integration region of $t_{n-1}$ requires the sequential pairwise ordering of integration variables.
That is, we must start with ordering the pair $t_{n-2}$, $t_{n-3}$, and then subsequently order the pair $t_{n-3}$, $t_{n-4}$ and so on until the pair $t_2$, $t_1$.
Each pairing will result in a nested factor of $1+e^{i\pi2\rho}$,
\be
\big(1+e^{i\pi2\rho}\big(1+e^{i\pi2\rho}\big(1+e^{i\pi2\rho}\big(\cdots\big)\big)\big)\big) = \sum_{m=0}^{n-1} e^{2\pi i\, m\rho}
\ee
It follows that the factor between the unordered integral $J_A$ and the ordered integral $I_A$ is:
\be
 \left(\sum_{m=0}^{n-1} e^{2\pi i\, m\rho} \right)\prod_{k=1}^{n-2} \sum_{m=0}^k e^{2\pi i\,m\rho} = \prod_{k=1}^{n-1} \sum_{m=0}^k e^{2\pi i\,m\rho} 
\ee
This can be simplified, since the sum is just a geometric progression.
\be
\prod_{k=1}^{n-1} \sum_{m=0}^k e^{2\pi i\,m\rho} = \prod_{k=1}^{n-1} \frac{e^{i\pi 2\rho (k+1)} - 1}{e^{i\pi 2\rho} - 1} = \prod_{k=1}^{n-1} e^{i \pi k \rho}\,\frac{\mf s\big((k+1)\rho\big)}{\mf s(\rho)}
\ee
This proves the general relation between $J_0$ and $I_0$. 

Finally, $J_A$ (for $0<A<n$) has two types of integration contours, with $A$ in one region and $n-A$ in another. Applying the above manipulations to this situation, one finds the relation between $J_A(\lambda)$ and $I_A(\lambda)$ (for arbitrary $A$) to be:
\be
  J_A (\lambda) = \left(\prod_{k=1}^{A-1} e^{i\pi k\rho}\, \frac{\mf s((k+1)\rho)}{\mf s(\rho)} \prod_{l=1}^{n-A-1} e^{i\pi l \rho}\,\frac{\mf s((l+1)\rho)}{\mf s(\rho)} \right)I_A(\lambda)
\ee
This proves \eref{JtoI}. 

\section{Degeneracies of quasi-characters}

\label{deg.2}

In this Appendix we have worked out the degeneracies of the first few quasi-characters in each of the four series Lee-Yang, $A_1, A_2, D_4$. The results are as follows:

      \begin{table}[h!]
        \centering
        \begin{tabular}{ccccccc}
          \toprule
          $r$ & $a=\frac{r}{5}-\frac{3}{10}$ & $c=\frac{2}{5}(6r+1)$ & $P=\frac{D_1}{D_0}$ & $D_0$ & $D_1$ & Remark \\
          \midrule
          0 & $-\frac{3}{10}$ & $\frac{2}{5}$ & 1 & 1 & 1 & Lee-Yang minimal model \vspace{0.1cm} \\
          1 & $-\frac{1}{10}$ & $\frac{14}{5}$ & 7 & 1 & 7 & $G_{2,1}$ WZW model \vspace{0.1cm} \\
          2 & $\frac{1}{10}$ & $\frac{26}{5}$ & 26 & 1 & 26 & $F_{4,1}$ WZW model \vspace{0.1cm} \\
          3 & $\frac{3}{10}$ & $\frac{38}{5}$ & 57 & 1 & 57 & $E_{7.5,1}$ WZW model \vspace{0.1cm} \\
          5 & $\frac{7}{10}$ & $\frac{62}{5}$ & 682 & 1 & 683 & Quasi-character \vspace{0.1cm} \\
          6 & $\frac{9}{10}$ & $\frac{74}{5}$ & 3774 & 1 & 3774 & Quasi-character \vspace{0.1cm} \\
          \bottomrule
        \end{tabular}
            \caption{The Lee-Yang series (Multiplicity  $M=1$)}
        \label{LYSeries}
      \end{table}
      
      \begin{table}[h!]
        \centering
        \begin{tabular}{ccccccc}
          \toprule
          $r$ & $a=\frac{r}{2}-\frac{1}{4}$ & $c=6r+1$ & $P=\frac{D_1}{D_0}$ & $D_0$ & $D_1$ & Remark \\
          \midrule
          0 & $-\frac{1}{4}$ & 1 & 2 & 1 & 2 & $SU(2)_1$ WZW Model \vspace{0.1cm} \\
          1 & $\frac{1}{4}$ & 7 & 56 & 1 & 56 & $E_{7,1}$ WZW model \vspace{0.1cm} \\
          2 & $\frac{3}{4}$ & 13 & 1248 & 1 & 1248 & Quasi-character \vspace{0.1cm} \\
          3 & $\frac{5}{4}$ & 19 & 26752 & 1 & 26752 & Quasi-character \vspace{0.1cm} \\
          4 & $\frac{7}{4}$ & 25 & 565760 & 1 & 565760 & Quasi-character \vspace{0.1cm} \\
          5 & $\frac{9}{4}$ & 31 & $\frac{83232768}{7}$ & 7 & 83232768 & Quasi-character \vspace{0.1cm} \\
          \bottomrule
        \end{tabular}
                \caption{The $A_1$ series (Multiplicity $M=1$)}
        \label{A1Series}
      \end{table}
      
      \begin{table}[h!]
        \centering
        \begin{tabular}{ccccccc}
          \toprule
          $r$ & $a=\frac{r}{3}-\frac{1}{6}$ & $c=4r+2$ & $P=\frac{D_1}{D_0}$ & $D_0$ & $D_1$ & Remark \\
          \midrule
          0 & $-\frac{1}{6}$ & 2 & 3 & 1 & 3 & $SU(3)_1$ WZW Model \vspace{0.1cm} \\
          1 & $\frac{1}{6}$ & 6 & 27 & 1 & 27 & $E_{6,1}$ WZW model \vspace{0.1cm} \\
          3 & $\frac{5}{6}$ & 14 & 1701 & 1 & 1701 & Quasi-character \vspace{0.1cm} \\
          4 & $\frac{7}{6}$ & 18 & 13122 & 1 & 13122 & Quasi-character \vspace{0.1cm} \\
          6 & $\frac{11}{6}$ & 26 & 767637 & 1 & 767637 & Quasi-character \vspace{0.1cm} \\
          7 & $\frac{13}{6}$ & 30 & 5845851 & 1 & 5845851 & Quasi-character \vspace{0.1cm} \\
          \bottomrule
        \end{tabular}
                \caption{The $A_2$ series (Multiplicity $M=2$)}
        \label{A2Series}
      \end{table}
      
      \begin{table}[h!]
        \centering
        \begin{tabular}{ccccccc}
          \toprule
          $r$ & $a=r$ & $c=12r+4$ & $P=\frac{D_1}{D_0}$ & $D_0$ & $D_1$ & Remark \\
          \midrule
          0 & 0 & 4 & 8 & 1 & 8 & $SO(8)_1$ WZW Model \vspace{0.1cm} \\
          1 & 1 & 16 & 4096 & 1 & 4096 & Quasi-character \vspace{0.1cm} \\
          2 & 2 & 28 & 1835008 & 1 & 1835008 & Quasi-character \vspace{0.1cm} \\
          3 & 3 & 40 & 805306368 & 1 & 805306368 & Quasi-character \vspace{0.1cm} \\
          4 & 4 & 52 & $\frac{2456721293312}{7}$ & 7 & 2456721293312 & Quasi-character \vspace{0.1cm} \\
          5 & 5 & 64 & $\frac{457396837154816}{3}$ & 3 & 457396837154816 & Quasi-character \vspace{0.1cm} \\
          \bottomrule
        \end{tabular}
                \caption{The $D_4$ series (Multiplicity $M=3$)}
        \label{D4Series}
      \end{table}

\newpage

\bibliographystyle{JHEP}

\bibliography{contour}

\end{document}